\documentclass[fleqn,usenatbib]{mnras}

\usepackage{newtxtext,newtxmath}

\usepackage[T1]{fontenc}

\DeclareRobustCommand{\VAN}[3]{#2}
\let\VANthebibliography\thebibliography
\def\thebibliography{\DeclareRobustCommand{\VAN}[3]{##3}\VANthebibliography}

\usepackage{graphicx}	
\usepackage{amsmath}	
\usepackage{amssymb}	
\usepackage{siunitx}
\usepackage{comment}
\usepackage{multirow}
\usepackage{physics}

\title[Full spectral-timing model]
{A full spectral-timing model to map the accretion flow in black hole binaries: the low/hard state of MAXI J1820+070}

\author[T. Kawamura et al.]{
Tenyo Kawamura,$^{1,2}$\thanks{E-mail: tenyo.kawamura@ipmu.jp}
Magnus Axelsson,$^{3,4}$
Chris Done$^{5,2}$
and Tadayuki Takahashi$^{2,1}$
\\
$^{1}$Department of Physics, University of Tokyo, Bunkyo, Tokyo 113-0033, Japan\\
$^{2}$Kavli Institute for the Physics and Mathematics of the Universe (WPI), University of Tokyo, Kashiwa, Chiba 277-8583, Japan\\
$^{3}$Oskar Klein Center for CosmoParticle Physics, Department of Physics, Stockholm University, SE-10691 Stockholm, Sweden\\
$^{4}$Department of Astronomy, Stockholm University, SE-10691 Stockholm, Sweden\\
$^{5}$Department of Physics, University of Durham, South Road, Durham DH1 3LE, UK
}

\date{Accepted 2022 January 1. Received 2021 November 15; in original form 2021 July 21}

\pubyear{2021}

\begin{document}
\label{firstpage}
\pagerange{\pageref{firstpage}--\pageref{lastpage}}
\maketitle


\begin{abstract}

The nature and geometry of the accretion flow in the low/hard state of black hole binaries is currently controversial. 
While most properties are generally explained in the truncated disc/hot inner flow model, the detection of a broad residual around the iron line 
argues for strong relativistic effects from an untruncated disc. 
Since spectral fitting alone is somewhat degenerate, we combine it with the additional information in the fast X-ray variability and perform a full spectral-timing analysis for {\it NICER} and {\it NuSTAR} data on a bright low/hard state of MAXI J1820+070.
We model the variability with propagating mass accretion rate fluctuations by combining two separate current insights: that the hot flow is spectrally inhomogeneous, and that there is a discontinuous jump in viscous time-scale between the hot flow and variable disc. 
Our model naturally gives the double-humped shape of the power spectra, and the increasing high-frequency variability with energy in the second hump. 
Including reflection and reprocessing from a disc truncated at a few tens of gravitational radii quantitatively reproduces the switch in the lag-frequency spectra, from hard lagging soft at low frequencies (propagation through the variable flow) to the soft lagging hard at the high frequencies (reverberation from the hard X-ray continuum illuminating the disc).
The viscous time-scale of the hot flow is derived from the model, and we show how this can be used to observationally test ideas about the origin of the jet. 
\end{abstract}


\begin{keywords}
accretion, accretion discs -- black hole physics -- X-rays: binaries
\end{keywords}



\section{Introduction}
\label{sec:1}

Spectral studies of the low/hard state in black hole binaries lead to conflicting conclusions about the nature and geometry of the accretion flow. 
The hard Comptonised continuum spectra seen in this state cannot arise from an isotropically emitting corona over the inner accretion disc (\citealt{Poutanen_2018} and references therein). 
A truncated disc/hot flow geometry is most often proposed to solve the energetic constraints, with the truncation radius decreasing for softer spectra as the source increases in brightness towards the transition to the high/soft (disc dominated) state (see e.g. \citealt{Done_2007}). 
However, this is at odds with more detailed spectroscopy around the iron line, where the line profile appears so broad as to require an untruncated disc (e.g. \citealt{Garcia_2015, Xu_2018, Garcia_2018}). The additional information encoded in the fast ($0.01$--$100\,\si{s}$) large amplitude variability which is characteristic of the low/hard state can potentially break the spectral controversy by giving an independent set of diagnostics into the nature and geometry of the state (e.g. \citealt{Uttley_2014}).

One key advantage of the truncated disc 
geometry is that it also gives a framework to explain the overall variability properties. 
A decreasing truncation radius between the disc and hot flow as the spectrum softens in the transition from the low/hard (Compton dominated) to high/soft (disc dominated) spectra
means a decreasing characteristic time-scale for all variability. 
This is seen in the obvious decrease in time-scale picked out by both the low-frequency break in the broad-band fast variability power spectrum and the 
prominent low-frequency ($0.1$--$10\,\si{Hz}$) quasi-periodic oscillation (QPO) seen in the Comptonised emission
(\citealt{Wijnands_1999}). 
Both these can be quantitatively explained in a model where the inner radius of the disc is the outer radius of the hot flow, with the low-frequency break in the broad-band variability set by fluctuations in mass accretion rate stirred up on the viscous time-scale of the outer flow, while the QPO is Lense-Thirring (relativistic vertical) precession
of the entire hot flow (\citealt{Ingram_2009, Ingram_2011, Ingram_2012}). 

There are other diagnostics of the fast time-scale variability which again can be most easily interpreted in the truncated disc/hot inner flow geometry. 
Combined spectral and timing constraints show that fluctuations in higher energy bands generally lag behind the same fluctuations seen in lower energy bands, and that the size of this lag decreases as the time-scale of the fluctuations decreases (\citealt{Miyamoto_1989, Revnivtsev_1999, Wilkinson_2009, Uttley_2014, Grinberg_2014}). 
Compton scattering should imprint a lag as a function of energy due to the additional light travel time required for each successive scattering order, but this is much
shorter than observed lags and has no dependence on fluctuation time-scale (\citealt{Miyamoto_1989, Nowak_1999}). Instead, the observed lags
are now generally interpreted as being related to the propagation time-scale of fluctuations through the accretion flow. Slow 
fluctuations are produced at the larger radii in the flow and have to propagate down 
through the entire flow to reach the innermost radii, so have the longest time lags. Conversely, faster fluctuations are produced at smaller radii, so have a shorter distance to travel to reach the innermost regions,  and hence shorter time lag (\citealt{Lyubarskii_1997, Kotov_2001, Arevalo_2006}). 
Propagation lags do not give any observable signature if the accretion flow is homogeneous, emitting the same spectrum at each radius. 
The additional key assumption required to explain the propagation lags in the data is that the flow is also radially stratified in its energy spectrum, with softer spectra emitted at larger radii, and harder spectra emitted closer to the black hole. This is a natural consequence of the truncated disc/hot inner flow geometry, as parts of the flow which are closest to the disc will see more seed photons for Compton cooling, so the spectrum will be softer than that close to the black hole, where fewer seed photons from the disc penetrate and where the energy released from gravity is highest (\citealt{Axelsson_2018}).

Radial stratification of the Comptonised emission also gives a possible solution to the dilemma of the iron line profile. A sum of softer and harder Comptonisation components gives a spectrum which is subtly curved.
Fitting a multi-temperature Comptonisation continuum with a single temperature model leaves broad residuals. 
The iron line modelling assumes that all these residuals arise from reflection on the disc,
but while the ionised reflection models do contain curvature, this is accompanied by  strong atomic line features. 
Only by dramatically smearing the lines by extreme relativistic effects can these be smoothed into a pseudo-continuum to match the curvature in the data (\citealt{Makishima_2008}, \citealt{Kohlemainen_2014}, \citealt{Basak_2017}, \citealt{Zdziarski_2021a}). 
The derived extreme gravity is then an artefact of continuum complexity . 

However, the iron line also gives an additional way to determine the inner disc edge. It 
is produced by the (fluctuating) hard X-rays illuminating the disc, so its emission must be lagged by the light travel time. This is difficult to see in the data as the iron line is a fairly small feature, and is produced at energies where standard CCD X-ray detectors are not so sensitive. However, X-ray illumination of the disc also produces a reflected continuum (Compton hump, peaking at 30~keV) and (quasi)thermalised emission from the X-ray heated disc photosphere, peaking below 1~keV. The latter can easily be studied with CCD instruments as long as the interstellar column to the source is low, and `soft lags' --- so called as the disc reprocessed emission is at soft energies and lags the hard X-ray continuum variability --- are now clearly seen (\citealt{Uttley_2011}). 
Their evolution to shorter time-scales as the source makes the transition from the low/hard to high/soft state supports the truncated disc models (\citealt{deMarco_2015, deMarco_2017, deMarco_2021}).

Despite the overall successes of the truncated disc/hot inner flow geometry, developing fully quantitative models for all these spectral-timing features has proved difficult.
On the spectral fitting alone, a complex continuum is not quite enough to make the truncated disc reflection models be the best statistical fit to the data (\citealt{Buisson_2019, Zdziarski_2021b}). Similarly, it is difficult to fully model the power spectra and the lags as a function of both frequency and energy (\citealt{Rapisarda_2016, Mahmoud_2018a, Mahmoud_2019}). 

Here we use {\it NICER} and {\it NuSTAR} data from the extremely bright recent outburst of MAXI J1820+070 (\citealt{Tucker_2018, Kawamuro_2018}) in its low/hard state. 
The inclination of the source has been measured as $66$--$81^\circ$ from the orbit (\citealt{Torres_2020}) and as $64 \pm 5^\circ$ from the jet (\citealt{Wood_2021}).
The source is relatively nearby (distance $\sim 3.0\,\si{kpc}$; \citealt{Atri_2020}), and has only small interstellar absorption along its line of sight (galactic absorption $\sim 10 ^{21}\,\si{cm^{-2}}$; \citealt{Uttley_2018}) so the spectrum can be studied down to the softest energies. {\it NICER} can handle the high count rates required for fast timing studies better than any other current detector, so the combination of object and instrument gives the best data so far on the disc in this controversial state. 

We develop a spectral-timing model including both the propagating mass accretion rate fluctuations and reverberation through a full
truncated disc/radially stratified hot inner flow geometry. 
Propagation and reverberation together play a key role in the fast variability, and various models combining them have been developed and applied to both AGN (\citealt{Wilkins_2013, Gardner_2014, Wilkins_2016}) and black hole binaries (\citealt{Rapisarda_2017a, Mahmoud_2019}).
The most self-consistent way of spectral-timing modelling is to simultaneously fit both the spectral and Fourier spectral properties. 
However, this is extremely expensive in CPU time
(\citealt{Mahmoud_2019}). 
Instead we build on earlier techniques which fix the spectral components from detailed fits to the time-averaged energy spectra, and then couple these to a timing model of the variability which includes both propagation and reverberation
(e.g. \citealt{Gardner_2014, Mahmoud_2018a}).
This approach gives a quantitative model which by definition fits all the major features of the energy spectra. 
Here we build a variability model using the energy spectral components and show that this can also broadly fit the power spectra as a function of energy and the lags seen between different energy bands as a function of frequency, including the switch between propagation (soft lead) and reverberation (soft lag).

While the model still has some limitations, we show that a truncated disc at a few tens of gravitational radii reproduces the overall features of the power spectra and lag-frequency spectra. 
We also show how the propagated variability allows us to map out the accretion speed as a function of radius in the hot flow. 
This depends on the efficiency of the underlying angular momentum transport processes, which depends on the magnetic field geometry and strength. 
Our results favour models where the transport is dominated by a small scale dynamo process (the magnetorotational instability: MRI \citealt{Balbus_1991, Balbus_1998}) giving Standard And Normal Evolution (SANE) rather than 
models with maximal magnetic flux (Magnetically Arrested Disc: MAD), where the dynamo is suppressed and accretion occurs via instabilities in the large scale magnetic fields (\citealt{Narayan_2012}). Thus spectral-timing models hold out the possibility of 
observationally testing models linking the jet to the hot flow. 


\section{Observation and Data Reduction}
\label{sec:2}

We study the bright low/hard state of the black hole binary MAXI J1820+070. 
We choose to use data from 2018-03-21, close to the peak in high energy flux, where there are simultaneous {\it NICER} and {\it NuSTAR} data (\citealt{Kara_2019, Buisson_2019, Zdziarski_2021b}). These data are summarised in Table \ref{tab:obs_info}.

The {\it NICER} data were processed using the {\it NICER} data analysis software, \verb'NICERDAS'  2019-06-19\_V006a in  \verb'HEASoft' version 6.26, and \verb'CALDB' 20200722.
The unfiltered data were cleaned with the standard calibration and standard time screening through \verb'nicercal' and \verb'nimaketime' (\citealt{Stevens_2018}).
Then the data were barycentre-corrected with the tool \verb'barycorr'.
Following reported caveats by the {\it NICER} team, we excluded the occasionally noisy focal plane modules 14 and 34 from the created cleaned event file.

The energy spectrum and light curves were extracted with \verb'XSELECT'. We excluded periods of
high background by discarding episodes where the $13$--$15\,\si{keV}$ light curve exceeds $2\,\si{counts/s}$
(\citealt{Ludlam_2018}).
The power spectra, cross spectra, and time lags as a function of frequency were calculated from light curves in segments of 65.536~s with 1 ms time bins ($2^{16}$ points) and rebinned logarithmically with the constant factor of 1.2 by following the prescriptions described in \cite{Uttley_2014} and \cite{Ingram_2019}.
Any segments with data gaps were discarded, which yields the smaller net exposure for timing analysis of $2294\,\si{s}$.

The {\it NuSTAR} data were reduced with the \verb'NuSTARDAS' pipeline v.1.8.0 and \verb'CALDB' 20200912.
To filter passages through the South Atlantic Anomaly, we set \verb'saamode=strict' and \verb'tentacle=yes'.
Following reported caveats by the {\it NuSTAR} team, we used  \verb'statusexpr="STATUS==b0000xxx00xxxx000"' to remove spurious triggers caused by bright sources.
The source and background regions were extracted with the circle of $60''$ radius, in which the former was centred on the peak brightness and the latter was placed in the lowest apparent source contamination. 
The source is so bright in this observation that the background is negligible.
However, the {\it NuSTAR} count rate is not high enough to give strong constraints on the high-frequency timing behaviour which we focus on here, so we use these data only to better define the spectral model. 
We use only data from a single detector module, FPMA, to illustrate the high-energy spectral shape and to avoid the cross calibration between the FPMA and FPMB modules. 

\begin{table}
 \caption{Basic information about the observations used in the paper from 2018-03-21.}
 \label{tab:obs_info}
 \begin{tabular}{p{30pt}p{40pt}p{40pt}p{40pt}p{40pt}}
  \hline
  Project & Start time &End time & Obs. ID & Exposure (s)\\
  \hline
  {\it NICER}  & 09:15:20  &23:15:40& 1200120106  & 5365\\
  {\it NuSTAR} & 07:06:09  &16:31:09& 90401309006 & 3539\\
  \hline
 \end{tabular}
\end{table}


\section{Spectral-timing models}
\label{sec:3}

A full spectral-timing model requires that we specify both the spectrum {\it and} its variability as a function of radius, with the propagation of fluctuations and reverberation through this radial structure. 

Previous attempts at this have used a variety of different assumptions tailored to the observed bandpass. 
\cite{Ingram_2012} modelled the timing properties assuming 
the spectrum was a homogeneous hot flow (single Comptonisation), with emissivity assumed to be a power law as a function of radius. The variability in this flow was assumed to be generated 
and propagated on the local viscous time-scale, which depends on radius.
This picture can be linked to models of angular momentum transport in the flow. The MRI in particular causes fluctuations in all physical quantities, and fluctuations in density propagate inwards on the local viscous time-scale,
producing fluctuations in mass accretion rate, which affects the observed luminosity (\citealt{Lyubarskii_1997, Kotov_2001, Arevalo_2006}).
This model could fit to {\it RXTE} data, where the low-energy bandpass limit of $>3$~keV meant that the disc was not important spectrally. However, propagated disc variability is likely important in the hot flow 
even when not in the observed bandpass (\citealt{Uttley_2011}). Not including a distinct low-frequency variability component from the disc
in the hot flow variability
means the model struggles to reproduce the breadth of the power spectrum, requiring uncomfortably large intrinsic variability power and spectral emissivity at very small radii \citep{Ingram_2011, Ingram_2012, Ingram_2013}. This remains a problem even when the Comptonisation region is radially stratified, with softer Comptonisation continua produced at larger radii (\citealt{Mahmoud_2018a}).

Instead, \cite{Rapisarda_2016} show that including fluctuations from the disc
can relieve these tensions, where the disc 
time-scale for fluctuations is discontinuously lower than the viscous time-scales in the hot flow at the disc truncation radius, consistent with physical expectations of the lower scale height disc structure compared to the hot flow. Importantly, this naturally gives the double-humped shape of the power spectra,
which are otherwise very difficult to reproduce (\citealt{Mahmoud_2018b}). Including lower-frequency variability from the disc also means that the hot flow needs only to produce the higher-frequency variability, which reduces the emissivity and variability requirements at small radii which are otherwise quite stringent (\citealt{Mahmoud_2018b}).

However, the \cite{Rapisarda_2016} models 
did not include full spectral information for the hot flow. 
Instead, they assumed the hot flow emissivity was a power law as a function of radius, and fit for the 
emissivity in each energy band. 
This allowed them to fit power spectra as a function of energy, and to calculate the lags between two energy bands, but does not use all the information contained in the energy spectrum. 

\cite{Mahmoud_2018a} used the energy spectrum to define the intrinsic spectral components (disc, soft and hard Comptonisation), together with the reflected spectra. 
The intrinsic spectra were assumed to be strictly radially stratified in order to couple these with a variability model (generation and propagation of fluctuations) to make a full spectral-timing description of the data. 
They extended the formalism in \cite{Mahmoud_2018b, Mahmoud_2019},
but the bumpiness of their derived power spectra was predominantly due to the enhancement of variability/emissivity at specific radii.

Here, we revisit the approach of \cite{Mahmoud_2018a}, fitting for the spectral components and then using these to define a full spectral-timing model.
But we couple this to the timing model incorporating the insight of \cite{Rapisarda_2016} where the disc time-scale is discontinuously lower than that in the hot flow. 


\section{Fitting the spectral components}
\label{sec:4}

\begin{figure}
\includegraphics[width=0.9\columnwidth]{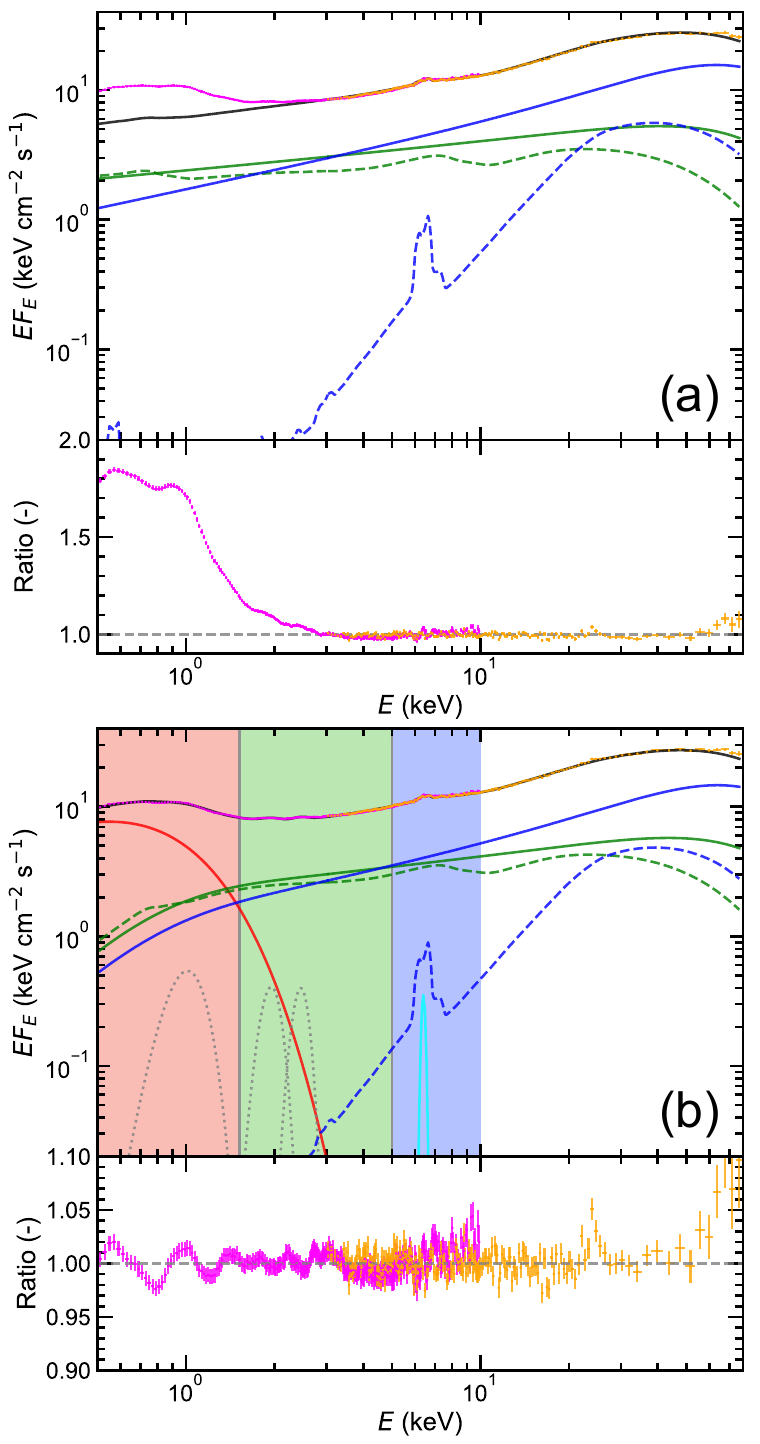}
    \caption{Sequence of fits to {\it NICER} (magenta) and {\it NuSTAR} (orange) spectra. 
    The upper and lower panels show the unabsorbed spectra and data-to-model ratios, respectively.
    The energy spectra are normalised to the {\it NuSTAR} effective area.
    (a) The {\it NuSTAR} spectrum alone is fitted with two Comptonisation components and reflection of these components from the truncated disc. 
    The black solid line shows the total spectrum.
    The hard (soft) Comptonisation $S_{\mathrm{hc}} (E)$ ($S_{\mathrm{sc}} (E)$) and its reflection $S_{\mathrm{hr}} (E)$ ($S_{\mathrm{sr}} (E)$) are shown with blue (green) solid and dashed lines.
    (b) Both the {\it NICER} and {\it NuSTAR} spectra are fitted with the addition of a disc component $S_{\mathrm{d}} (E)$ (red).
    The unphysical low-energy cutoff of the Comptonisation and reflection components are corrected.
    A narrow Gaussian component centred at 6.4~keV (cyan) is included.
    In addition, three Gaussian components around 1.0, 1.9, and 2.4~keV (grey, dotted) are included to compensate for residuals.
    The derived parameters are summarised in Table. \ref{tab:ni1200120106_eeuf_rat}.
    The shaded energy bands, low (red), mid (green), and high (blue), are mainly used in the timing analysis.}
	\label{fig:ni1200120106_eeuf_rat}
\end{figure}

The spectral fits were performed with the X-ray spectral fitting package 
{\tt XSPEC} (\citealt{Arnaud_1996}). 
We use the
energy ranges $0.5$--$10\,\si{keV}$ and $3$--$78\,\si{keV}$ for the {\it NICER} and {\it NuSTAR} data, respectively. 
Following the recommendation by the {\it NICER} team to handle calibration uncertainties\footnote{\url{https://heasarc.gsfc.nasa.gov/docs/nicer/analysis_threads/cal-recommend/}}, we add 1\,\% systematic errors to the {\it NICER} data.
In addition, we allow the relative normalisation of {\it NICER} with respect to {\it NuSTAR} to be a free parameter, but assume the same spectral shape for both instruments. 

Previous detailed spectral analyses of the {\it NuSTAR} data have shown that fits with a thermal disc and hot Comptonisation have some tensions, even when including two reflection components of different ionisation and typical radii.
Both \cite{Zdziarski_2021b} and \cite{Buisson_2019} show that this modelling leads to best-fit solutions where one of the reflectors is strongly smeared, requiring $R_{\mathrm{in}}\sim 2R_{\mathrm{g}}$, completely inconsistent with a truncated disc. 
Furthermore, there are other tensions in this fit, with the reflectors having surprisingly high Fe overabundance of $Z_{\mathrm{Fe}} \sim 6$ and surprisingly low inclination of $\sim 30^\circ$. 
Adding a second continuum component relieves these tensions, with iron overabundance reduced to $Z_{\mathrm{Fe}} \sim 2$ for the measured binary and jet inclination of $\sim 60$--$65^\circ$. 
Each continuum component is associated with its own reflection component (now with different shapes due to the differences in the spectral index as well as the different ionisation). 
With this spectral complexity, the derived inner disc radius increases to $\sim 100~R_{\mathrm{g}}$, consistent with a truncated disc. Nonetheless, the derived solution is still in the highly relativistic regime as the softer continuum lamppost source has a height of $\sim 2~R_{\mathrm{g}}$ (\citealt{Zdziarski_2021b}).

We use the spectral model of \cite{Zdziarski_2021b} (our data correspond to their epoch 2) with their two Comptonisation continua and reflection components as our baseline, i.e., we use the {\tt compps} model continuum and corresponding {\tt reflkerrG} ionised reflection models. 
These are required when fitting to the {\it NuSTAR} data as the standard {\tt nthcomp} thermal Comptonisation models in {\tt XSPEC} do not give the proper shape of the high-energy rollover (\citealt{Niedzwiecki_2019, Zdziarski_2020}). 
This impacts the reflected continuum as well, so the {\tt xillverCp} reflection models also underestimate the high-energy continuum. 

We fit first to the {\it NuSTAR} data alone and confirm that our fit matches that of \cite{Zdziarski_2021b}. 
We then extend the model down to the {\it NICER} energy range and see clear evidence for excess thermal emission from a disc component, as shown in Fig. \ref{fig:ni1200120106_eeuf_rat} (a). 
We include this with a {\tt diskbb} model, a simple sum of blackbody components, characterised by a peak temperature from the inner edge of the disc, $kT_{\mathrm{bb}}$. 

However, the {\tt reflkerrG} reflection model, which 
is used to define the direct Comptonisation spectrum as well as the reflected spectrum, fixes the seed photon of the direct Comptonisation spectrum at a blackbody temperature of $1\,\si{eV}$\footnote{\url{https://users.camk.edu.pl/mitsza/reflkerr/reflkerr.pdf}}.
In addition, the reflection models do not include the seed photon temperature of the incident Comptonisation as a free parameter, and the reflection spectra are computed for a fixed seed photon temperature equivalent to a disc blackbody temperature of $50\,\si{eV}$\footnote{\url{http://www.sternwarte.uni-erlangen.de/~dauser/research/relxill/}}.
As a result, whereas the seed photons at $kT_{\mathrm{bb}}\sim 0.22\,\si{keV}$ are obtained from the spectral fitting, both the direct and reflected emission extend to lower energies than the seed photons in an unphysical way. 
This likely has an impact on attempts to measure the disc density from the low-energy extent of the reflected emission (e.g. \citealt{Jiang_2019}).

\begin{table}
\caption{Model parameters for the spectral fit to the {\it NICER} and {\it NuSTAR} data.
 Errors represents 90\,\% confidence intervals.
 The seed photon temperature for the soft and hard Comptonisation components is tied to the disc temperature $kT_{\mathrm{bb}}$ via the local model {\tt xslowe}. 
 Fixed parameters are 
 the galactic absorption $N_{\mathrm{H}}=1.4 \times 10^{21}\,\si{cm^{-2}}$, 
 the electron temperature $kT_{\mathrm{e}}=23\,\si{keV}$,
 the peak energy of the narrow Gaussian component of $6.4\,\si{keV}$, 
 the black hole spin $a^{*}=0$, 
 the inner radii of the reflection region of $R_{\mathrm{in}}=14, 170 R_{\mathrm{g}}$ for the soft and hard Comptonisation components, 
 the outer radii of the reflection region of $R_{\mathrm{out}}=1000 R_{\mathrm{g}}$ for both Comptonisation components, 
 the inclination of $66^{\circ}$, 
 the Fe abundance of $Z_{\mathrm{Fe}}=1.1$, and 
 the ionisation parameter $\log _{10} \xi = 0.43$ for the reflection of the hard Comptonisation component.
 The photon index, normalisation, fraction of the reflection, and the standard deviation of the Gaussian component are denoted as $\Gamma$, $\mathrm{norm}$, $\mathcal{R}$, and $\Delta E$, respectively.
 The reduced $\chi^2$ is $2725.5/2805$.}
 \label{tab:ni1200120106_eeuf_rat}
 \begin{tabular}{p{70pt}p{40pt}p{40pt}p{50pt}}
  \hline
  Component           &Model&Parameter                               & Value\\
  \hline \hline
  Disc                &\texttt{diskbb}&$kT_{\mathrm{bb}}$ (keV)                  &$0.23 \pm 0.001$\\[2pt]
  (red in Fig. \ref{fig:ni1200120106_eeuf_rat} (b))&&$\mathrm{norm}$ ($10^{5}$)                &$3.90 \pm 0.91$ \\[2pt]
  \hline
  Soft Comptonisation &\texttt{reflkerrG}&$\Gamma$                                  &$1.74 ^{+0.02} _{-0.01}$ \\[2pt]
  and reflection (green) &&$\log _{10} \xi$                          &$3.30 ^{+0.01} _{-0.02}$\\[2pt]
               &&$\mathrm{norm}$                           &$2.29 ^{+0.16} _{-0.29}$ \\[2pt]
                      &&$\mathcal{R}$                             &$0.96 ^{+0.12} _{-0.17}$ \\[2pt]
  \hline
  Hard Comptonisation &\texttt{reflkerrG}&$\Gamma$                                  &$1.45 \pm 0.01$ \\[2pt]
  and reflection (blue) &&$\mathrm{norm}$                           &$1.53 \pm 0.21$ \\[2pt]
                &&$\mathcal{R}$                             &$0.54 ^{+0.13} _{-0.08}$ \\[2pt]
  \hline
  Gaussian (cyan)            &\texttt{gauss}&$\Delta E$ (keV)                          &$0.10 ^{+0.08} _{-0.10}$ \\[2pt]
                &&$\mathrm{norm}$ ($10^{-3}$)               &$2.21 ^{+0.58} _{-0.49}$ \\[2pt]
  \hline
 \end{tabular}
\end{table}

In order to correct for the seed photon break, we make a local 
{\tt XSPEC} model, {\tt xslowe}, which multiplies both the reflected and direct Comptonisation spectra by 
the difference between the low-energy Comptonisation rollover for any given seed photon temperature and that assumed in the {\tt relxill} and {\tt reflkerr} models$^{2, 3}$ (see also \citealt{Garcia_2014, Garcia_2016}).
However, we stress that this is only an approximate correction factor. The reflection models of \cite{Garcia_2016} conserve energy, so the energy removed by suppressing the unphysical low-energy extent must emerge somewhere, most likely peaking around $3$--$4kT_{\mathrm{bb}}$ like a blackbody or disc blackbody, i.e., around 1~keV. This lack of reprocessed emission in our best-fitting model becomes important when we consider the reverberation signal (see Section \ref{sec:7_3}).

The residuals are dominated by a clear atomic feature at 1~keV from a combination of Ne\,X and Fe\,L shell transitions. A $\sim$~5~\% feature is seen at 1~keV in the best calculations of the disc photosphere at similar temperatures to those here of $0.22$~keV
(\citealt{Davis_2006}). Not all elements are ionised at these temperatures, so full radiative transfer through solar abundance material is required to calculate these features. 
However, these models are tabulated including relativistic corrections, assuming that the disc extends down to the innermost stable circular orbit. Thus they can be used to fit disc dominated states 
(e.g. LMC X-3: \citealt{Kubota_2010}), but the relativistic corrections are not appropriate for a truncated disc as assumed here. 
Hence we cannot use these photosphere models to directly fit to our data, and therefore just add a Gaussian component at $\sim 1$~keV. 

The residuals then show a smaller set of features at  $1.5$--$2.5\,\si{keV}$, probably from highly ionised Si and S. These are not produced significantly in the photosphere models, so are 
more likely explained by the 
limitations in current reflection models discussed above. 
Again, we compensate for these using Gaussian components
at $1.9$ and $2.4\,\si{keV}$, respectively. It is important to model these features as otherwise the higher statistics of the {\it NICER} data pull the fit away from the {\it NuSTAR}.

There is then also a small residual 
around the iron line, so we additionally include a narrow Gaussian component to model this emission. 
This results in $\Delta\chi^2=152$ for the addition of two free parameters, so is highly significant. 
The resulting line EW is $\sim 8$~eV, similar to that seen in a similar bright low/hard state of GX339-4 (\citealt{Done_2010}).
We suggest that this is from the strong optical wind which is seen in this state (\citealt{Munoz-Darias_2019, Sai_2021}), and stress that this should be included in all detailed models of the iron line profile (see discussion in \citealt{Axelsson_2021}). 

In summary, our employed model is written as 
\\
{\tt tbabs*(diskbb+xslowe*reflkerrG(SC)\\+xslowe*reflkerrG(HC)+gauss+gauss+gauss+gauss)}.
\\
The unabsorbed data and model components, all normalised to the {\it NuSTAR} effective area, are shown in Fig. \ref{fig:ni1200120106_eeuf_rat} (b).
The galactic absorption is taken into account with {\tt tbabs}.
The blackbody component emitted from the truncated disc is calculated by {\tt diskbb}.
The Comptonisation components and relevant reflection components shown with solid and dashed lines in green (soft: SC) or blue (hard: HC) colour are calculated with {\tt xslowe*reflkerrG}.
The three Gaussian components to correct for residuals below 3~keV
are shown in grey dotted lines, while the cyan solid line compensates for the additional narrow 
core iron line emission at $6.4\,\si{keV}$.
The model parameters are tabulated in Table \ref{tab:ni1200120106_eeuf_rat}. 
The figure also shows the three energy bands which we will use for the power spectra, low (red, $0.5$--$1.5\,\si{keV}$), 
mid (green, $1.5$--$5.0\,\si{keV}$) and high (blue, $5.0$--$10.0\,\si{keV}$). 

Clearly (and model independently), the low energy band has mainly disc emission with some Comptonisation, while the high energy band has mainly Comptonisation and very little disc emission. 
Thus the low and high energy band power spectra should be distinctly different as they are dominated by quite different components. 
In our particular spectral model, the soft and hard Comptonisation components contribute about equally to the mid and high energy bands, which means their power spectra should be quite similar. We note that other spectral decompositions (e.g. a soft Comptonisation component which is more like a hotter blackbody component, as derived e.g. by \citealt{Dzielak_2021} for these data or \citealt{DiSalvo_2001} for Cygnus X-1) would predict instead that the mid and high energy bands were dominated by different components so could have distinct variability. 

We stress that none of the variability results presented in Section \ref{sec:5} depend on the detailed shape of the model components.
It becomes important only in Section \ref{sec:7} where the detailed timing properties are computed using this spectral decomposition.


\section{Variability Data}
\label{sec:5}

\subsection{Power spectra}
\label{sec:5_1}
\begin{figure}
	\includegraphics[width=0.9\columnwidth]{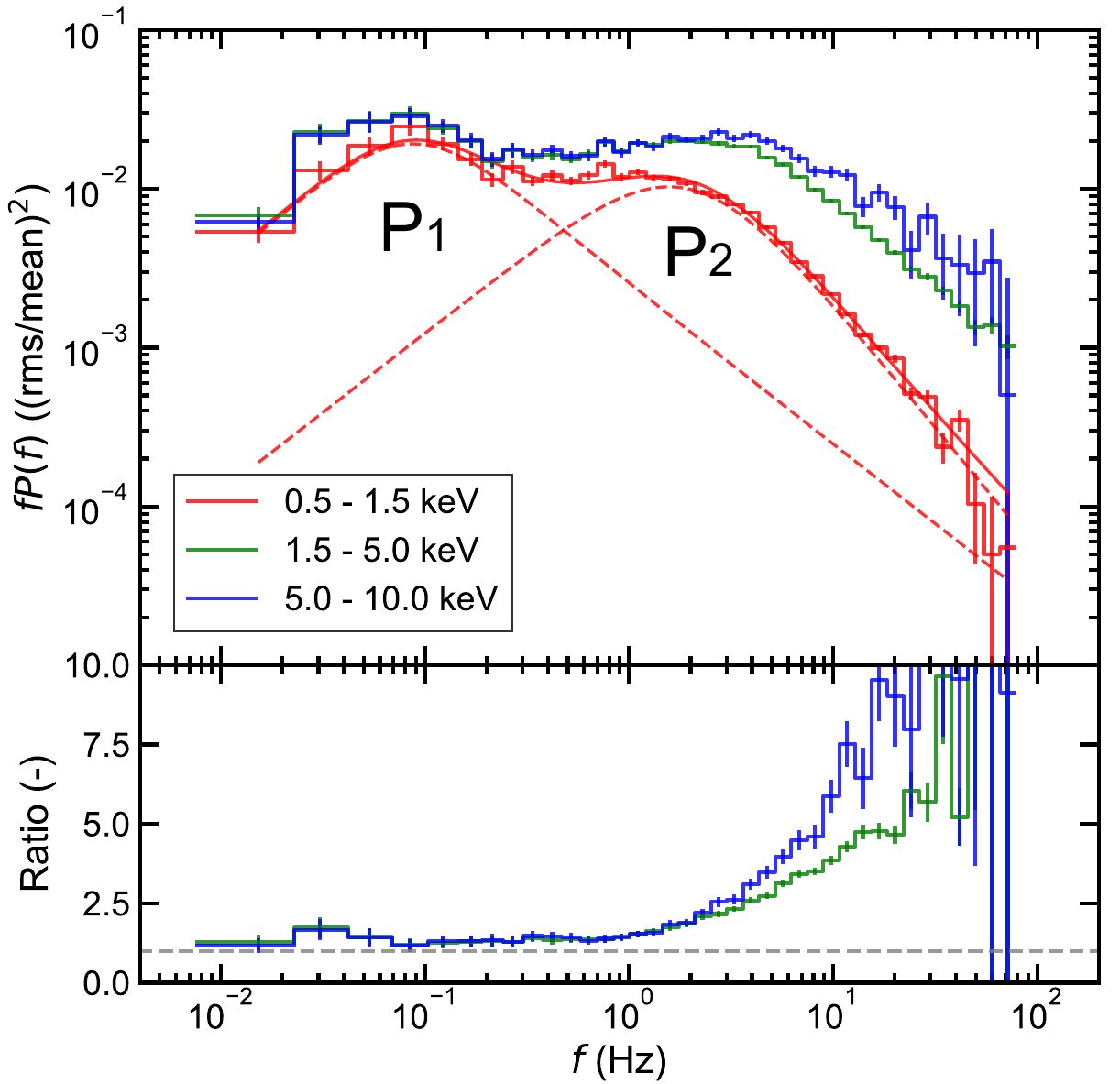}
    \caption{Upper panel: power spectra in three energy bands calculated from the {\it NICER} data.
    The red smooth lines show the fit to the power with the sum (solid) of two broad-band variability components $\mathrm{P}_{1}$ and $\mathrm{P}_{2}$ (dashed). 
    Lower panel: ratio of the power spectra of the mid (green) and high (blue) energy bands compared to the low energy band.
    In both panels, the vertical error bars represent $1\sigma$.}
	\label{fig:ni1200120106_psd}
\end{figure}

Variability on a wide range of time-scales can be examined by power spectra.
We calculate them from the {\it NICER} data, subtracting the Poissonian noise contribution (\citealt{Uttley_2014}) and adopting the normalisation such that the integral of the power spectra corresponds to the fractional variance (\citealt{Belloni_1990, Miyamoto_1991, vanderKlis_1997, Vaughan_2003}).
The power spectra in three energy bands, low ($0.5$--$1.5\,\si{keV}$, red), mid ($1.5$--$5.0\,\si{keV}$, green), and high ($5.0$--$10.0\,\si{keV}$, blue), are shown in the upper panel of Fig.~\ref{fig:ni1200120106_psd}. 
It is seen that they can be characterised by two broad-band variability components.
We refer to the slower and faster components as $\mathrm{P}_{1}$ and $\mathrm{P}_{2}$, respectively.

The lower panel of Fig.~\ref{fig:ni1200120106_psd} highlights the much stronger high-frequency variability at higher energies by showing the ratio of the power spectra in the mid and high energy bands with respect to that in the low energy band. 
However, it also shows a very 
small increase in variability at low frequencies, around $0.03$--$0.04$~Hz. 
This is primarily due to a weak QPO with fundamental around $0.036\,\si{Hz}$ seen at this time, together with its second harmonic at $0.072\,\si{Hz}$ (\citealt{Buisson_2019, Ma_2021}). 
The observed low-frequency variability in any energy band is then the sum of the broad-band noise $\mathrm{P}_{1}$ and the QPOs.
The QPOs are predominantly seen in the Comptonisation components so increase the ratio of the mid/high to low-energy power spectra at the QPO frequencies. Without the contribution from the QPO, 
we see that the underlying low-frequency component $\mathrm{P}_{1}$ is similar in shape and normalisation in all energy bands, peaking at $0.08\,\si{Hz}$.

By contrast, the energy-dependent behaviour of $\mathrm{P}_{2}$ appears to be intrinsic. 
It has long been known that the higher energy bands have more variability in short time-scales, this is generally modelled in data above $3\,\si{keV}$ as having the same shape but higher normalisation.
The current data are the first to enable us to deeply explore the behaviour of $\mathrm{P}_{2}$ at lower energy bands, where the thermal emission from the disc dominates the energy spectrum. 
Including the low energy band makes it clear that the frequency at which the $\mathrm{P}_{2}$ peak occurs is changing as well as its normalisation.
We study the energy dependence of $\mathrm{P}_{2}$ in more detail below.

\subsection{Modelling the power spectra to characterise the shift in peak frequency of $\mathrm{P}_2$ with energy}
\label{sec:5_2}
We now use power spectra over narrower energy bands to look at the behaviour of $\mathrm{P}_{2}$ in more detail. 
We fit these power spectra with the sum of the two broad-band variability components $\mathrm{P}_{1}$, $\mathrm{P}_{2}$. 
Note that neglecting the weak QPOs does not affect the estimates of the peak frequency of $\mathrm{P}_{2}$.

The low-frequency component $\mathrm{P}_{1}$ can be fit with a standard (symmetric) Lorentzian 
\begin{equation}
    P_{1}(f)=A_{1} \frac{(\Delta f_{1})^2}{(f-f_{\mathrm{c}1})^2 + (\Delta f_{1})^2},
    \label{eq:p1}
\end{equation}
where $A_{1}$ is the normalisation, $f_{\mathrm{c1}}$ the centroid frequency, and $\Delta f_{1}$ the half width at the half maximum (HWHM).
On the other hand, the high-frequency component $\mathrm{P}_{2}$ is clearly asymmetric. 
Hence, we include an additional parameter $\beta_2$ to control the 
relative shape of the low- and high-frequency wing of $\mathrm{P}_{2}$. 
Assuming further that this Lorentizan is zero-centred, we model $\mathrm{P}_{2}$ with  
\begin{equation}
    P_{2}(f)=A_{2} \left( \frac{(\Delta f_{2})^2}{f^2 + (\Delta f_{2})^2} \right) ^{\beta_{2}}. 
    \label{eq:p2}
\end{equation} 
Thus $fP_2(f)$ tends to $A_2 f$ for $f\ll \Delta f$ like the standard zero-centred Lorentzian, whereas at high frequencies it is proportional to $f^{-2\beta_2+1}$
rather than $f^{-1}$, making it asymmetric on a $\log f - \log fP(f)$ plot for $\beta_2 \neq 1$.
A sample fit to the broader energy band power spectrum is shown overlaid on the upper panel in Fig.~\ref{fig:ni1200120106_psd}.

\begin{figure}
	\includegraphics[width=0.85\columnwidth]{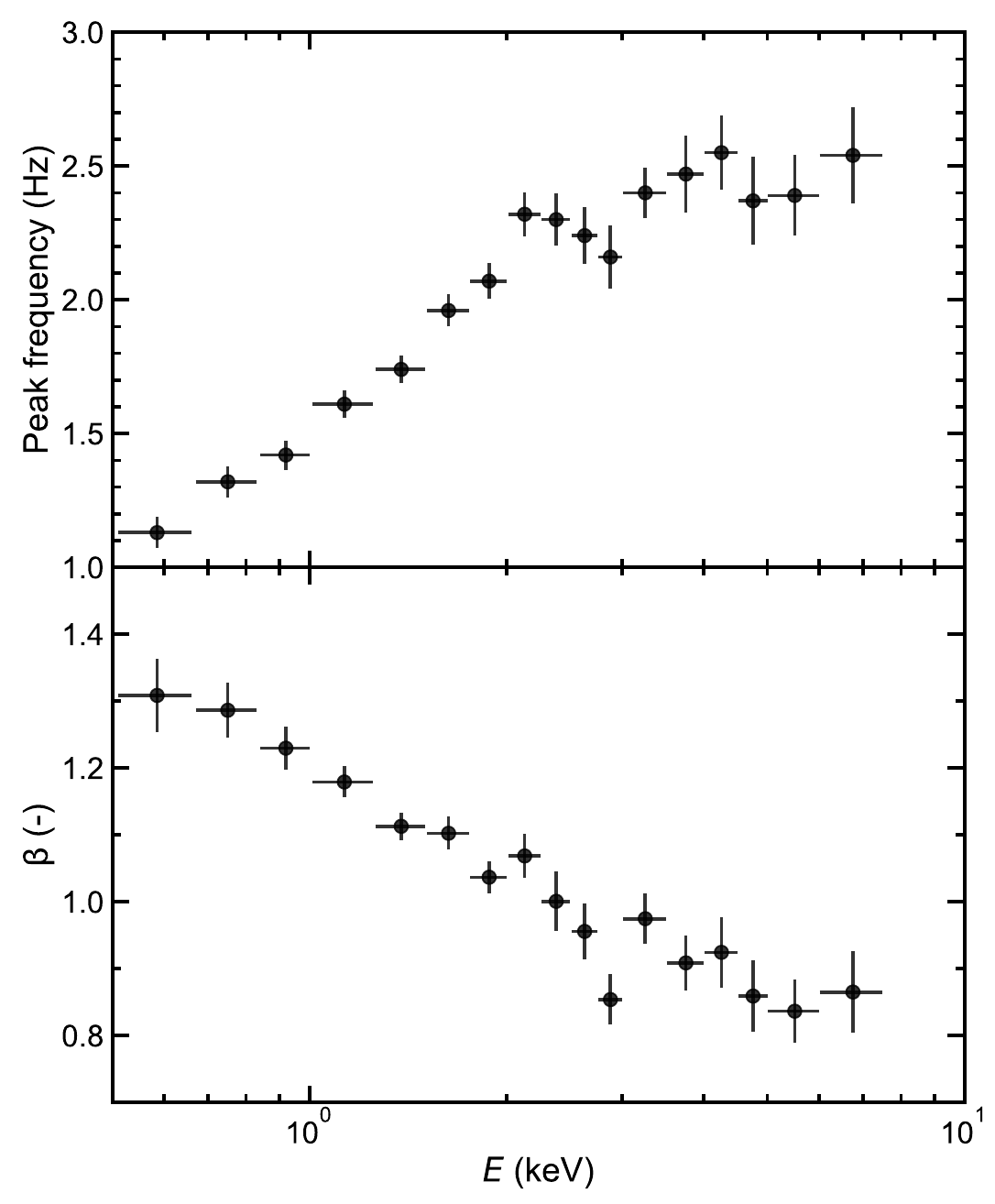}
    \caption{Energy dependence of the peak frequency (upper) and the index of the asymmetric Lorentzian $\beta _2$ (lower) for the higher-frequency variability component $\mathrm{P}_{2}$.
    The vertical error bars represent 68\,\% confidence intervals derived from the Monte Carlo technique.
    The data in $7.5$--$10.0\,\si{keV}$ band are removed due to large uncertainties.}
	\label{fig:ni1200120106_nu2_vs_ene}
\end{figure}

To obtain the energy dependence of the peak frequency, we perform the fit to the power spectrum in each energy band with the sum of two variability components, $P_{1}(f)+P_{2}(f)$. 
For $\mathrm{P}_{1}$ we fix the peak frequency, $f_{\mathrm{c1}}$, and width, $\Delta f_{1}$, as we find that these are almost constant for different energy bands.
The free parameters left are thus $A_{1}$, $A_{2}$, $\Delta f_{2}$, and $\beta_{2}$. We show the energy dependence of the peak frequency as well as index $\beta_2$ governing the asymmetry of  $\mathrm{P}_{2}$ is shown in Fig.~\ref{fig:ni1200120106_nu2_vs_ene}, where the 
errors are derived by a Monte Carlo technique (\citealt{Press_1992}) from 1000 simulated power spectra.

It is clear that the majority of the shift in peak frequency of $\mathrm{P}_{2}$ with energy occurs
below $2\,\si{keV}$.
This corresponds to the energy at which the spectrum makes a transition from being dominated by the thermal disc to the Comptonisation (see Fig.~\ref{fig:ni1200120106_eeuf_rat}). 
Thus this shift in peak frequency should be related to the transition  between the disc and hot flow. 
By contrast, $\beta_2$ shows a more constant decline with energy,
but crosses unity (standard zero centred Lorentian) at around 2~keV. 
At higher energies $\beta_{2}$ is less than unity, meaning there is more high-frequency power than in a standard zero-centred Lorentian.
At lower energies, the index $\beta_{2}$ is greater than unity, so as well as the decrease in peak frequency of $\mathrm{P}_{2}$, there is a sharper drop in power spectrum at frequencies beyond the peak (see low-energy power spectrum, red in Fig.~\ref{fig:ni1200120106_psd}).
This is the first time that these properties have been seen in such detail. 
They show a shift in behaviour at the energy where the spectrum shows the transition between emission from the disc and hot flow, so they give us a new way to explore what is happening in this controversial region. 

\subsection{Lag-frequency spectrum}
\label{sec:5_3}

\begin{figure}
	\includegraphics[width=0.9\columnwidth]{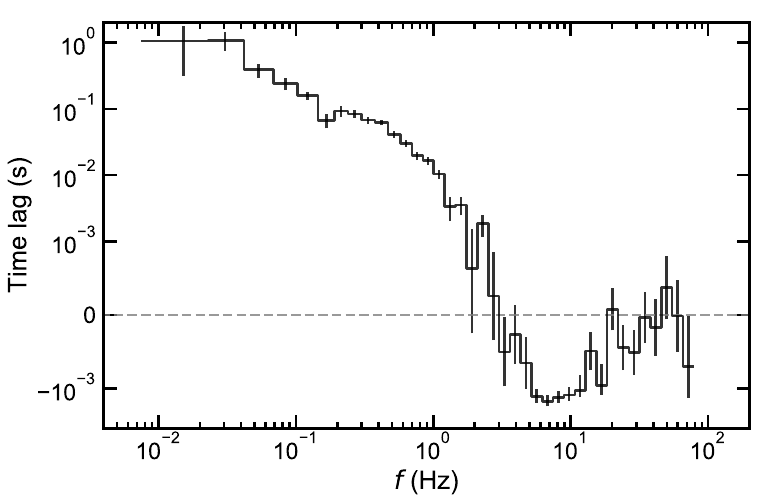}
    \caption{Lag-frequency spectra between the soft band $0.5$--$1.5\,\si{keV}$ and hard band $2.0$--$10.0\,\si{keV}$.
    The positive time lag means fluctuations in the hard band lag behind those in the soft band and vice versa.
    The symmetric logarithmic scale is employed for the representation of the time lags , i.e., linear between $-10^{-3}\,\si{s}$ and $10^{-3}\,\si{s}$ and logarithmic otherwise.
    The vertical error bars represent $1\sigma$.
    The grey dashed line shows the time lags of zero.}
	\label{fig:lagf_obs}
\end{figure}

Another important aspect of the variability is time lags as a function of frequency as these show the causal connection between different spectral components.
The lag-frequency spectrum between the soft band ($0.5$--$1.5\,\si{keV}$) and a total hard band ($2.0$--$10.0\,\si{keV}$, extended down to $2$~keV to maximise counts) is shown in Fig.~\ref{fig:lagf_obs}.
The soft bands is primarily related to the disc while the hard is dominated by Comptonisation. 
Positive time lags mean variations in hard photons lag behind those in soft photons, whereas negative time lags mean variations in soft photons lag behind those in hard photons. 

At longer time-scales, the hard band lags the soft band.
On the other hand, the causal connection switches around $\sim 3\,\si{Hz}$, giving rise to the soft band lagging the hard band at shorter time-scales. Such
low-frequency hard lags and high-frequency soft lags are typically observed in black hole binaries (\citealt{Uttley_2011, Kara_2019}). 
They are explained by propagating mass accretion rate fluctuations to give the hard lags (e.g. \citealt{Kotov_2001}) and thermal reverberation of the Comptonised emission illuminating the disc making the soft lags.
In the following sections, we quantitatively model the lag-frequency spectra as well as power spectra by combining propagation and reverberation through the radially stratified spectral components.

\section{Variability model formalism}
\label{sec:6}

Our model builds on the analytic scheme proposed by \citealt{Ingram_2013}, in which the power spectrum and cross spectrum of the local mass accretion rate are calculated first and converted into those of the flux.
Thus, we mainly focus on aspects of prescription that differ from those in \citealt{Ingram_2013}.
The full technical method to calculate the power spectra and time lags of the flux is described in Appendix~\ref{sec:app_a}, but we outline the various steps below.

\subsection{Accretion geometry and spectral stratification}
\label{sec:6_1}

We assume that there are three spectral components, 
a thermal component from the disc, $S_{\mathrm{d}}(E)$, together with soft, $S_{\mathrm{sc}}(E)$, 
and hard, $S_{\mathrm{hc}}(E)$, Comptonisation components from the hot flow.
Modelling the hot flow with two Comptonisation components comes from the spectral decomposition performed in Section~\ref{sec:4}.
It is also supported by the fast variability since the hard time lags typically observed above 2 keV are difficult to explain with a single Comptonisation component (e.g., \citealt{Mahmoud_2018a} for Cygnus X-1).
We set the time-averaged shapes of these 
from the results of the spectral fit 
(Fig.~\ref{fig:ni1200120106_eeuf_rat} (b), Table~\ref{tab:ni1200120106_eeuf_rat}) 
All these $S$ are photon flux spectra, since the important quantity for variability is the number of photons. 

\begin{figure}
	\includegraphics[width=0.95\columnwidth]{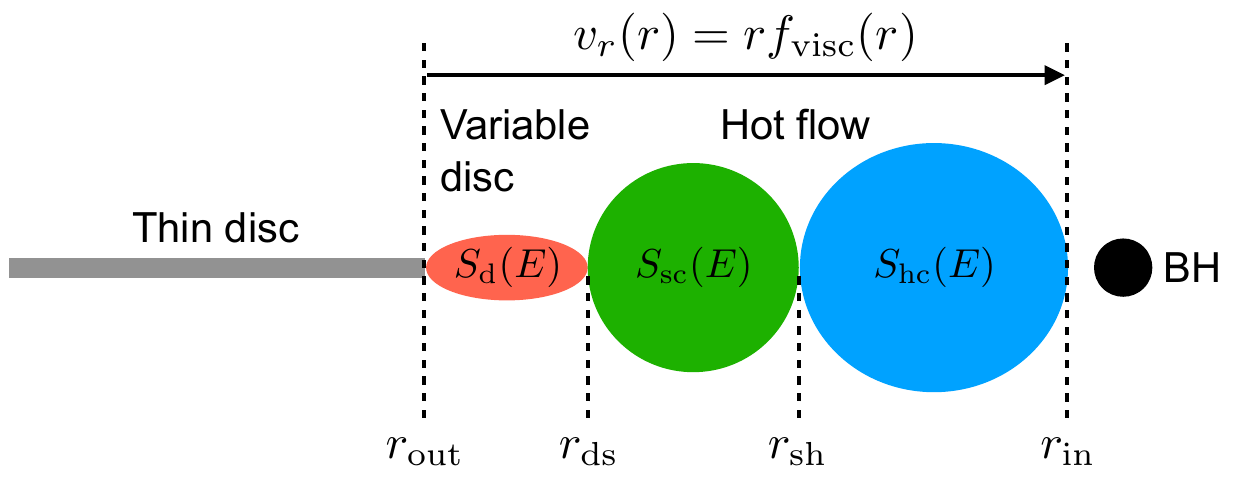}
    \caption{Assumed geometry in the model. 
    The emission regions of the three spectral components, that is the thermal component $S_{\mathrm{d}}(E)$ and two Comptonisation components $S_{\mathrm{sc}}(E)$, $S_{\mathrm{hc}}(E)$, are radially stratified.
    Each region is spectrally homogeneous. 
    The variable flow is located between $r_{\mathrm{in}}$ and $r_{\mathrm{out}}$, whereas the thin disc located further does not contribute to the variability.
    The local fluctuations propagate inwards throughout the variable flow with the speed of the radial velocity $v_{r}(r)$, which is determined by the radius from the central object, $r$, and viscous frequency $f_{\mathrm{visc}} (r)$.}
	\label{fig:model_geometry}
\end{figure}

We also assume that these spectral components are radially stratified, 
where the variable disc region extends from 
$r_{\mathrm{out}}$ down to $r_{\mathrm{ds}}$, 
the soft Comptonisation component from $r_{\mathrm{ds}}$ to $r_{\mathrm{sh}}$, and 
the hard Comptonisation component from $r_{\mathrm{sh}}$ to $r_{\mathrm{in}}$, as shown in Fig. \ref{fig:model_geometry}.
In this geometry, we refer to the truncation radius as the outer radius of the variable disc $r_{\mathrm{out}}$.
All radii expressed with the small letter, $r$, are given in units of the gravitational radius $R_{\mathrm{g}}=GM/c^2$.
We fix $M=8\,M_{\odot}$ (\citealt{Torres_2019}).

The edges of the flow $r_{\mathrm{in}}$ and $r_{\mathrm{out}}$ are model parameters, whereas the transition radii $r_{\mathrm{sh}}$ and $r_{\mathrm{ds}}$ are calculated from the energy dissipation (\citealt{Mahmoud_2019})
\begin{equation}
    \begin{split}
    &\frac{\int dE\,ES_{\mathrm{d}} (E) }{\int dE\,ES_{\mathrm{hc}} (E)}=\frac{\int _{r_{\mathrm{ds}}} ^{r_{\mathrm{out}}} dr\,2\pi r \epsilon (r) }{\int _{r_{\mathrm{in}}} ^{r_{\mathrm{sh}}} dr\,2\pi r \epsilon (r)}, \\
    &\frac{\int dE\,ES_{\mathrm{sc}} (E) }{\int dE\,ES_{\mathrm{hc}} (E)}=\frac{\int _{r_{\mathrm{sh}}} ^{r_{\mathrm{ds}}} dr\,2\pi r \epsilon (r) }{\int _{r_{\mathrm{in}}} ^{r_{\mathrm{sh}}} dr\,2\pi r \epsilon (r)},
    \end{split}
\label{eq:trans_radii}
\end{equation}
where $\epsilon (r)$ is the emissivity in the variable flow.
Since it remains unclear, we simply assume the emissivity of the standard disc (\citealt{Shakura_1973, Novikov_1973, Pringle_1981}) 
\begin{equation}
\epsilon (r) \propto  r^{-\gamma} b(r),\,\, \gamma=3,\,\, b(r)=1-\sqrt{r_{\mathrm{in}}/r}
\end{equation}
for the whole variable flow throughout the paper.
We note that taking the thermal reverberation into account in Section \ref{sec:7_3} modifies the top equation of  (\ref{eq:trans_radii}) (see Appendix~\ref{sec:app_b}).

\subsection{Mass accretion rate fluctuations}
\label{sec:6_2}

We divide the variable flow between $r_{\mathrm{in}}$ and $r_{\mathrm{out}}$ into $N_{\mathrm{r}}$ rings logarithmically such that $\Delta r_{n}/r_{n}$ is constant, 
where $r_{n}$ and $\Delta r_{n}$ ($n=1, \cdots, N_{\mathrm{r}}$ from outer to inner rings) are the central radius of the $n$th ring measured from the central object and the distance between central radii of neighbouring rings. 
The number of rings $N_{\mathrm{r}}$ affects the resolution of the model.

The intrinsic fluctuations of the local mass accretion rate at the $n$th ring, $a(r_{n}, t)$ (see Appendix~\ref{sec:app_a}), are assumed to be represented by the power spectrum with
a zero-centred Lorentzian function (\citealt{Ingram_2016}) 
\begin{equation}
    P(r_{n}, f)=\frac{2 \sigma ^2}{\pi \mu ^2} \frac{f_{\mathrm{visc}} (r_{n}) }{f^2 + (f_{\mathrm{visc}} (r_{n}))^2}\,\,\,\,(n=1, 2, \cdots, N_{\mathrm{r}}),
    \label{eq:mdot_psd_intrinsic}
\end{equation}
where $\mu$ and $\sigma ^2$ are the average and variance of $\dot{a}(r_{n}, t)$ and thus $P(r_{n}, f)$ is normalised such that $\int _{0} ^{+\infty} df\,P(r_{n}, f)=(\sigma/\mu)^2$. 
The HWHM of the Lorentzian is set by the local viscous frequency $f_{\mathrm{visc}}(r_{n})$, which is the reciprocal of the local viscous time-scale, $f_{\mathrm{visc}}(r)=1/t_{\mathrm{visc}}(r)$.
We set the average to unity, $\mu = 1$, and use normalised and dimensionless mass accretion rate.
We set the variance to $\sigma ^2 = \mu ^2 F_{\mathrm{var}} / N_{\mathrm{dec}}$, where $N_{\mathrm{dec}}$ is the number of rings per radial decade and thus $F_{\mathrm{var}}$ is the fractional variability per radial decade.
This prescription of the variability of the mass accretion rate assumes the same variability amplitude between the logarithmically spaced rings, which is often obtained from MRI simulations (\citealt{Beckwith_2008}).

The fluctuations of the local mass accretion rate propagate inwards with the radial velocity (accretion speed) written as (\citealt{Ingram_2016})
\begin{equation}
    v_{r} (r) = \frac{r}{t_{\mathrm{visc}}(r)} = r f_{\mathrm{visc}}(r) .
    \label{eq:radial_velocity}
\end{equation}
We note that the radial velocity (\ref{eq:radial_velocity}) is expressed in units of the speed of light $c$ when the viscous time-scale (frequency) $t_{\mathrm{visc}}(r)$ ($f_{\mathrm{visc}}(r)$) is expressed in units of $R_{\mathrm{g}}/c$ ($c/R_{\mathrm{g}}$).
Under this condition, the power spectrum for each ring and cross spectrum for each combination of two rings 
are calculated analytically (see Appendix~\ref{sec:app_a}).
In equations (\ref{eq:mdot_psd_intrinsic}) and (\ref{eq:radial_velocity}), 
the viscous frequency $f_{\mathrm{visc}}(r)$ determines both the characteristic frequency at which the fluctuation at a given radius is generated and the radial velocity through the flow. 
This corresponds to the physical picture that the variability faster than the viscous time-scales is attenuated through the diffusion and that the mass accretion happens in the viscous time-scales (\citealt{Churazov_2001, Ingram_2016}).

\begin{table*}
 \caption{Parameter values of the model used in the paper. 
 The meaning of the parameters is described in the text. 
 The transition radii $r_{\mathrm{sh}}$, $r_{\mathrm{ds}}$ are not free parameters but determined by equation (\ref{eq:trans_radii}) or its correction due to the implementation of the reprocessed emission.
 The sign ``--'' denotes that a parameter is unused.  
 Other parameters concerning the resolution of the model are set to $N_{\mathrm{r}}=100$, $\Delta t=5\,\si{ms}$, $N=2^{14}$ in common.}
 \label{tab:model_par}
 \begin{tabular}{p{20pt}p{20pt}p{20pt}p{20pt}p{20pt}p{20pt}p{20pt}p{20pt}p{20pt}p{20pt}p{20pt}p{20pt}p{20pt}}
  \hline
  Figure & $r_{\mathrm{in}}$ & $r_{\mathrm{sh}}$& $r_{\mathrm{ds}}$& $r_{\mathrm{out}}$ & $B_{\mathrm{d}}$ & $m_{\mathrm{d}}$ & $B_{\mathrm{f}}$ & $m_{\mathrm{f}}$&$F_{\mathrm{var}}$&$f_{\mathrm{rep}}$&$t_{0}$&$\Delta t_{0}$\\
  \hline
  \ref{fig:psd_comp_continuous_fvisc}        &6   &16.5&27.6&45&45&1.65&$=B_{\mathrm{d}}$&$=m_{\mathrm{d}}$&0.55&--&--&--  \\
  \ref{fig:psd_comp_discontinuous_fvisc}     &6   &16.5 &27.6&45&0.03&0.5&6.0&1.2&0.8&--&--&--  \\
  \ref{fig:psd_lagf_comp_discontinuous_fvisc}&6   &17.8 &32.1&45&0.03&0.5&6.0&1.2&0.8&0.4&$4.3\,\si{ms}$&$5.0\,\si{ms}$\\
  \hline
 \end{tabular}
\end{table*}

\subsection{Variability of the flux}
\label{sec:6_3}

The flux in the number of photons at energy $E$ can be written as the sum of the contributions from each ring:
\begin{equation}
    x(E, t)=\sum _{n=1} ^{N_{\mathrm{r}}} w (r_{n}, E) \dot{m}(r_{n}, t),
    \label{eq:flux_e_narrow}
\end{equation}
where the relation that the radiation energy at each ring $r_{n}$ is proportional to the local mass accretion rate $\dot{m}(r_{n}, t)$ is used.
The weight $w(r_{n}, E)$ regulates the contribution from the $n$th ring to the flux at the energy $E$ and is expressed as 
\begin{equation}
    w(r_{n}, E)=S(E) \frac{\epsilon (r_{n}) 2 \pi r_{n} \Delta r_{n}}{\sum _{j\,(r_{\mathrm{min}} < r_{j} \leq r_{\mathrm{max}} )} \epsilon (r_{j}) 2 \pi r_{j} \Delta r_{j}},
    \label{eq:weight_e_narrow}
\end{equation}
where 
$(S(E), r_{\mathrm{min}}, r_{\mathrm{max}})=(S_{\mathrm{hc}}(E), r_{\mathrm{in}},  r_{\mathrm{sh}})$, $(S_{\mathrm{sc}}(E), r_{\mathrm{sh}},\,r_{\mathrm{ds}})$, $(S_{\mathrm{d}}(E), r_{\mathrm{ds}},\,r_{\mathrm{out}})$ for $r_{\mathrm{in}} < r_{n} \leq r_{\mathrm{sh}}$, $r_{\mathrm{sh}} < r_{n} \leq r_{\mathrm{ds}}$, $r_{\mathrm{ds}} < r_{n} \leq r_{\mathrm{out}}$ respectively, depending on which spectral component the $n$th ring belongs to.
From equations (\ref{eq:flux_e_narrow}) and (\ref{eq:weight_e_narrow}), the time-averaged flux is reduced to $\langle x(E, t) \rangle =S_{\mathrm{hc}}(E)+S_{\mathrm{sc}}(E)+S_{\mathrm{d}}(E)$.
Expressing the flux with the form of equation (\ref{eq:flux_e_narrow}) is essential to calculate the power spectrum and cross spectrum of the flux analytically (see Appendix~\ref{sec:app_a}).

In equations (\ref{eq:flux_e_narrow}) and (\ref{eq:weight_e_narrow}), we assumed that $x(E, t)$, $w(r_{n}, E)$ are designated by a single energy $E$. 
However, in terms of the comparison with observations, we should take wide energy bands enough for the observations to have a high signal-to-noise ratio. 
The wide energy band can be incorporated simply by averaging the photon flux:
\begin{equation}
\begin{split}
    &w (r_{n}, (E_{\mathrm{min}}, E_{\mathrm{max}})) \\
    &=S(E_{\mathrm{min}}, E_{\mathrm{max}})
    \frac{\epsilon (r_{n}) 2 \pi r_{n} \Delta r_{n}}{\sum _{j\,(r_{\mathrm{min}} < r_{j} \leq r_{\mathrm{max}} )} \epsilon (r_{j}) 2 \pi r_{j} \Delta r_{j}},
\end{split}
    \label{eq:weight_e_wide}
\end{equation}
where the energy band is designated by the combination of the lower and upper bounds, $(E_{\mathrm{min}}, E_{\mathrm{max}})$, and 
\begin{equation}
    S(E_{\mathrm{min}}, E_{\mathrm{max}})=\frac{1}{E_{\mathrm{max}}-E_{\mathrm{min}}}\int _{E_{\mathrm{min}}} ^{E_{\mathrm{max}}} \,dE S(E).
    \label{eq:average_flux}
\end{equation} 
The expression of the flux with the energy band $(E_{\mathrm{min}}, E_{\mathrm{max}})$ keeps the form of equation (\ref{eq:flux_e_narrow}):
\begin{equation}
    x((E_{\mathrm{min}}, E_{\mathrm{max}}), t)=\sum _{n=1} ^{N_{\mathrm{r}}} w (r_{n}, (E_{\mathrm{min}}, E_{\mathrm{max}})) \dot{m}(r_{n}, t).
    \label{eq:flux_e_wide}
\end{equation}
In addition, the detector effective area $A_{\mathrm{eff}} (E)$ and galactic absorption $N_{\mathrm{H}} (E)$ should be taken into account since the contribution to the observed flux is statistically different between different energy bins.
This is incorporated with the substitution of $S(E) A_{\mathrm{eff}} (E) \mathrm{e}^{-N_{\mathrm{H}}(E)\sigma _{\mathrm{T}}}$ for $S(E)$ in equation (\ref{eq:weight_e_narrow}) and (\ref{eq:average_flux}) (\citealt{Mahmoud_2018a}), where $\sigma _{\mathrm{T}}$ is the Thomson cross section.
This substitution alters the time-averaged flux:$\langle x(E, t) \rangle \rightarrow (S_{\mathrm{hc}}(E)+S_{\mathrm{sc}}(E)+S_{\mathrm{d}}(E))A_{\mathrm{eff}} (E) \mathrm{e}^{-N_{\mathrm{H}}(E)\sigma _{\mathrm{T}}}$. 
We calculate the power spectra and cross spectra of the flux for wide energy bands, using the expression of the flux (\ref{eq:flux_e_wide}).
The summary of the parameters used in the model is found in Table~\ref{tab:model_par}.


\section{Modelling the variability}
\label{sec:7}

\subsection{Matching the behaviour of $\mathrm{P}_{2}$}
\label{sec:7_1}
We first explore whether the propagating fluctuations scenario can indeed match the observed shift in power spectral shape of $\mathrm{P}_{2}$ with energy. We do not attempt to reproduce $\mathrm{P}_{1}$ in this subsection. 

We assume that the viscous frequency $f_{\mathrm{visc}}(r)$ is continuous and characterised by a single power law.
Following previous studies (\citealt{Ingram_2011, Mahmoud_2018a}), we express the viscous frequency as 
\begin{equation}
    f_{\mathrm{visc}}(r)=Br^{-m}f_{\mathrm{K}}(r) \propto r^{-(m+3/2)} \,\,\,\,(r_{\mathrm{in}} \leq r \leq r_{\mathrm{out}}),
\end{equation}
where $f_{\mathrm{K}}(r)=(1/2\pi)r^{-3/2}(c/R_{\mathrm{g}})$ is the Keplerian frequency.

\cite{Ingram_2011} set the parameters $B=0.03$ and $m=0.5$ 
from the relation between the QPO frequency
(assumed to be from Lense-Thirring precession)
and low-frequency break in the broad-band variability
(assuming this is set by fluctuations propagating from 
the outer edge of the hot flow). 
Here instead we allow $B$ and $m$ in the hot flow to be free parameters as we do not model the low-frequency break (which is from $\mathrm{P}_{1}$) in this subsection. 

\begin{figure}
	\includegraphics[width=0.9\columnwidth]{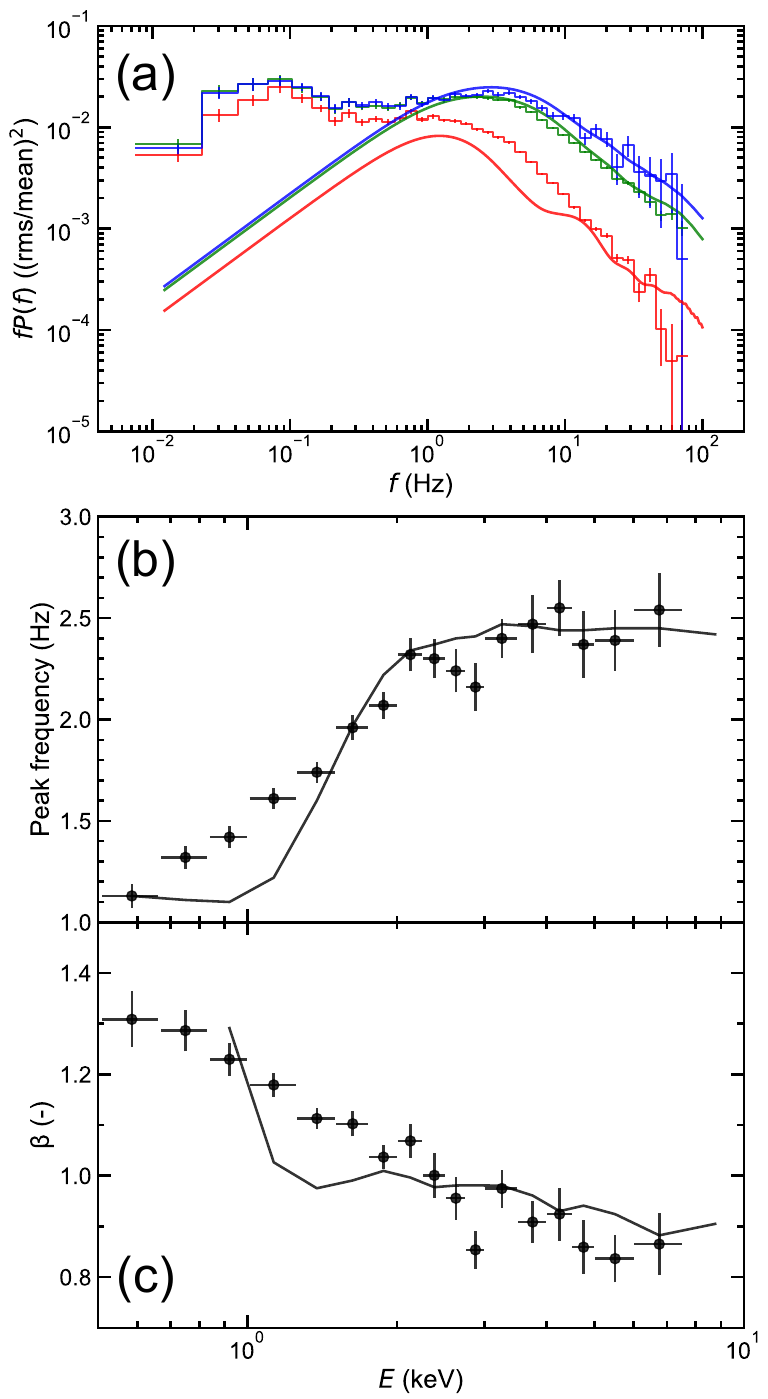}
    \caption{Comparison of power spectra (a) and the energy dependence of the peak frequency (b) and asymmetry (c) between the model (solid) and observation (stepped).
    Energy bands in the power spectra are the same as in Fig. \ref{fig:ni1200120106_psd}.
    The continuous viscous frequency throughout the variable flow is assumed in the model.
    Since the power spectra of low energy bands calculated from the model are bumpy at high frequencies, making the derivation of the index $\beta$ somewhat arbitrary, the indices at low energy bands are omitted.}
	\label{fig:psd_comp_continuous_fvisc}
\end{figure}

We assume $r_{\mathrm{in}}=6$ and set $r_{\mathrm{out}} =45$ for consistency with the next subsection where we discuss the origin of $\mathrm{P}_{1}$ in more detail.
The power spectra derived from a model with 
$B=45$, $m=1.65$ are shown in the upper panel of  Fig.~\ref{fig:psd_comp_continuous_fvisc}. Other model parameters are $F_{\mathrm{var}}=0.55$, $N_{\mathrm{r}}=100$, $\Delta t=5\,\si{ms}$, $N=2^{14}$, where $\Delta t$ is the sampling interval and $N$ the number of sampling points.
The peak frequency and index $\beta$ derived from 
the fit to the power spectra with the asymmetric Lorentzian (\ref{eq:p2}) are shown in the lower panels of Fig.~\ref{fig:psd_comp_continuous_fvisc}.

The energy-dependent power spectrum of $\mathrm{P}_{2}$ as well as its peak frequency and index $\beta$ are overall well reproduced.
Both the shift in peak frequency and the change in the asymmetry with respect to energy are produced by the different proportion of the three spectral components, i.e., the emission of higher energies comes from the regions closer to the central object, which have variability in higher frequencies.
The model particularly succeeds in reproducing the observed behaviour of $\mathrm{P}_{2}$ above $2\,\si{keV}$, where the emission from the hot inner flow dominates the total energy spectrum.
Thus, the propagating fluctuations scenario combined with the inhomogeneous hot flow explains well the behaviour of $\mathrm{P}_{2}$ above $2\,\si{keV}$, though it has a sharper
decrease below 2~keV than seen in the data. 

However, the much larger discrepancy is that the current picture completely ignores the lower-frequency variability which makes $\mathrm{P}_{1}$.
This is because the adjusted viscous frequency is fast compared to the prescription of \cite{Ingram_2011} ($B=0.03$, $m=0.5$).
If instead we change $B$ and $m$ to reproduce $\mathrm{P}_{1}$, it is not at all easy to reproduce $\mathrm{P}_{2}$ even with an extremely strongly centrally peaked emissivity and small inner radius (\citealt{Ingram_2011, Mahmoud_2018a}). 
Motivated by the work of \cite{Rapisarda_2016}, we then introduce a discontinuous jump in viscous frequency to reproduce both $\mathrm{P}_{1}$ and $\mathrm{P}_{2}$ simultaneously. 

\subsection{Discontinuity of the viscous time-scale between the disc and flow}
\label{sec:7_2}
We assume that the viscous frequency follows different power laws between the variable disc and hot flow.
It is included in our model as follows:
\begin{equation}
    f_{\mathrm{visc}}(r)=
    \begin{cases}
        B_{\mathrm{f}}r^{-m_{\mathrm{f}}} f_{\mathrm{K}}(r) & (r_{\mathrm{in}} \leq r \leq r_{\mathrm{ds}}),\\
        B_{\mathrm{d}}r^{-m_{\mathrm{d}}} f_{\mathrm{K}}(r) & (r_{\mathrm{ds}} \leq r \leq r_{\mathrm{out}}).
    \end{cases}
    \label{eq:fvisc_discont}
\end{equation}
The different viscous frequency prescriptions between the disc and hot flow are physically natural since the properties of the accretion flow are expected to be different between these regions.
The discontinuity of the viscous frequency at the transition radius $r_{\mathrm{ds}}$ produces two bumps (\citealt{Rapisarda_2016}). 
We expect these to directly correspond to the variability components $\mathrm{P}_{1}$ and $\mathrm{P}_{2}$ respectively.
Thus, we fix here $B_{\mathrm{d}}=0.03$ and $m_{\mathrm{d}}=0.5$ to preserve the QPO-low-frequency break relation (\citealt{Ingram_2011}). The observed low frequency break at $0.05$~Hz then fixes the inner radius of the disc at $45R_g$. 

\begin{figure}
	\includegraphics[width=0.9\columnwidth]{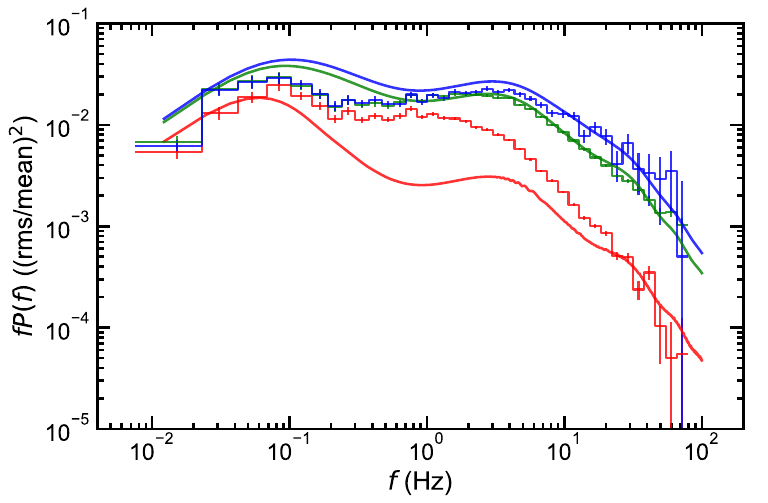}
    \caption{Power spectra obtained with a discontinuous viscous frequency prescription between the variable disc and hot flow. 
    Lines and colours are the same as in Fig. \ref{fig:psd_comp_continuous_fvisc} (a).}
	\label{fig:psd_comp_discontinuous_fvisc}
\end{figure}

The power spectra calculated with $B_{\mathrm{f}}=6.0$, $m_{\mathrm{f}}=1.2$, $F_{\mathrm{var}}=0.8$ and the same parameter values as in Fig.~\ref{fig:psd_comp_continuous_fvisc} are shown in Fig.~\ref{fig:psd_comp_discontinuous_fvisc}. 
The double-humped shape of the power spectra matches the observation, where $\mathrm{P}_{1}$ and $\mathrm{P}_{2}$ originate from the disc and hot flow, respectively.
The inwards propagation of fluctuations transmits $\mathrm{P}_{1}$ into the mid (green) and high (blue) energy bands, where the emission from the hot flow dominates the energy spectrum.

The clear defect of the model is that the low energy band (red) lacks variability in the $0.3$--$3\,\si{Hz}$ range. This is 
because it is dominated by the disc which only varies on the long time-scales of $\mathrm{P}_1$. There is a very small contribution from the soft and hard Comptonisation regions in this energy band, so there is a small contribution to higher frequency power from $\mathrm{P}_{2}$, but nowhere near enough to explain the data. 
We conclude that the power spectrum at energies where the disc emission is dominant cannot be explained by the propagating fluctuations process alone.

\begin{figure}
	\includegraphics[width=0.9\columnwidth]{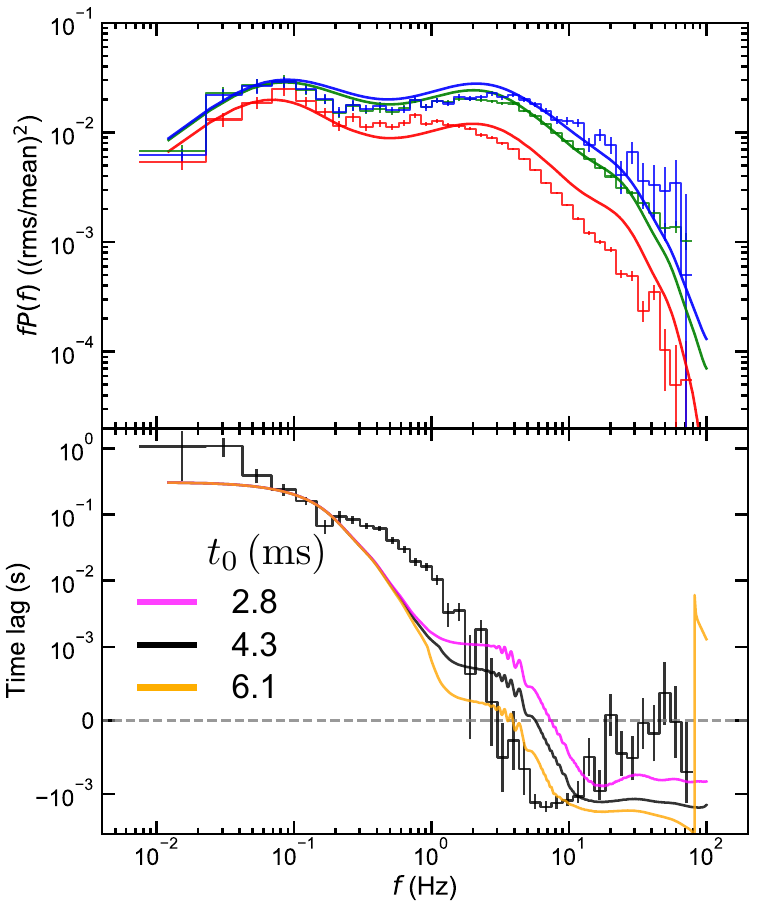}
    \caption{Power spectra (upper) and lag-frequency spectrum (lower) obtained with a discontinuous viscous frequency prescription and implementation of the reflected emission.
    Stepped and smooth lines show the observation and model, respectively.
    In the upper panel, colours are the same as in Fig. \ref{fig:psd_comp_continuous_fvisc}.
    In the lower panel, the energy bands are the same as in Fig. \ref{fig:lagf_obs}.
    For comparison, the lag-frequency spectra calculated with $t_{0}=2.8\,\si{ms}$ (magenta) and $t_{0}=6.1\,\si{ms}$ (orange) are shown. Larger time lags fit better to the maximum reverberation frequency, but exacerbate the discrepancy below 3~Hz.}
	\label{fig:psd_lagf_comp_discontinuous_fvisc}
\end{figure}

\begin{figure*}
	\includegraphics[width=0.95\linewidth]{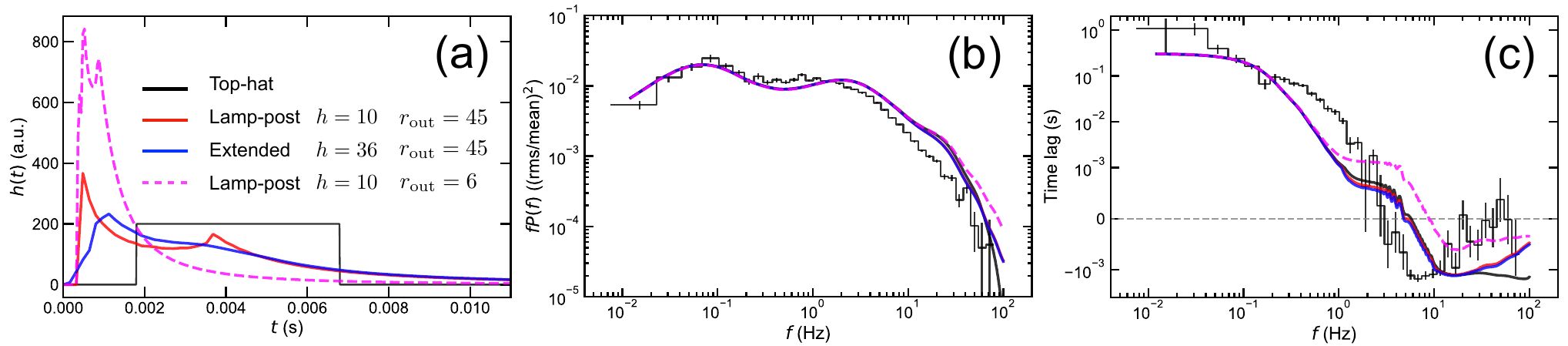}
    \caption{Comparison of impulse responses $h(t)$ (a), power spectra in the low energy band of $0.5$--$1.5\,\si{keV}$ (b), and lag-frequency spectra between $0.5$--$1.5\,\si{keV}$ and $2.0$--$10.0\,\si{keV}$ (c) for different formulations of the impulse response for reflection and reprocessing. 
    Parameter values other than $t_{0}$ and $\Delta t_{0}$ are common (see the bottom row of Table \ref{tab:model_par}).
    The black solid lines represent the top-hat impulse response.
    The red and blue solid lines are obtained with lamp-post and extended illuminating source geometries, respectively.
    The outer thin disc is assumed to be truncated in common.
    The magenta dashed line is calculated with the lamp-post source geometry and untruncated thin disc. 
    In (a), the impulse responses are normalised such that the integral over time in units of second becomes unity.
    In (b) and (c), the observation results are shown for comparison.}
	\label{fig:timeprop_diffgeo}
\end{figure*}

\subsection{Including reflection and reprocessing}
\label{sec:7_3}
The hot flow illuminates the disc, so the fast variability of the Comptonisation components is {\it reprocessed/reflected} by the disc. 
The process is required not only by the lack of faster variability component $\mathrm{P}_{2}$ in the low energy band in the power spectrum, but by the detection of the soft lag, which shows that some part of the low energy band emission lags behind the Comptonisation components, at high frequencies (see Fig.~\ref{fig:lagf_obs}).

Using only the reflected emission from the spectral fits, $S_{\mathrm{sr}}(E)$ and $S_{\mathrm{hr}}(E)$, does not give sufficient signal for (negative - soft lags hard) reverberation lags to ever dominate over the (positive - hard lags soft) propagation lags. 
This shows that there must be a larger contribution of reprocessing than is included in the current spectral model.
This is similar to the situation seen in the Narrow Line Seyfert 1 galaxy, PG~1211+026 (\citealt{Gardner_2014}). Here, the solution was to realise that the spectral models of ionised reflection are not always appropriate, especially where the seed photons for Comptonisation are 
not close to the assumed (hardwired) temperature of $50\,\si{eV}$
(see Section~\ref{sec:4}).

We include an additional reverberating component, assuming part of the disc component, $f_{\mathrm{rep}}S_{\mathrm{d}}(E)\,(0 \leq f_{\mathrm{rep}}<1)$, stems from thermalisation of the illuminating flux, not from direct disc emission.
Thus, the reflected spectrum employed in our model is $S_{\mathrm{sr}}(E)+S_{\mathrm{hr}}(E)+f_{\mathrm{rep}}S_{\mathrm{d}}(E)$.
Including the reflected component modifies the formulation of the model, which is summarised in Appendix~\ref{sec:app_b}.
We model how the disc responds to the 
illumination using a top-hat impulse response function 
\begin{equation}
    h(E, t)=
    \begin{cases}
        C(E) & (t_0 - \Delta t_0/2 < t \leq t_0 +\Delta t_0/2),\\
        0    & (\text{otherwise}),
    \end{cases}
\label{eq:impulse_response}
\end{equation}
where $t_0$ and $\Delta t_0$ represent the delay and duration, respectively.
The impulse response is different only in amplitude
between different energy bands. 
The factor $C(E)$ for infinitesimal energy bands (and more practically, $C(E_{\mathrm{min}}, E_{\mathrm{max}})$ for wide energy bands) is determined analytically from the conservation of the flux (see Appendix~\ref{sec:app_c}). 
A larger fraction of reprocessing means less 
intrinsic disc emission, which increases the inferred inner disc radius $r_{\mathrm{ds}}$ (see Table \ref{tab:model_par}).

There are four main free parameters which 
affect the power spectra and lag-frequency spectra, namely the average reverberation time lag, $t_{0}$, the fraction of reprocessed disc flux, $f_{\mathrm{rep}}$, and the radial dependence of the 
propagation speed in the hot flow, set by 
$B_{\mathrm{f}}$ and $m_{\mathrm{f}}$. The combination of these four parameters
which best encompasses all of the data is shown in 
Fig.~\ref{fig:psd_lagf_comp_discontinuous_fvisc}.
How the model changes with each parameter is shown in Appendix~\ref{sec:app_d}.

This best overall model reproduces both the power spectra as a function of energy and the lag-frequency spectrum fairly well. 
The reprocessed emission not only produces more 
variability at high frequencies in the low energy band as required, but also forms the soft lag at high frequencies.
The fraction of the reprocessed emission is set to $f_{\mathrm{rep}}=0.4$, which yields $r_{\mathrm{sh}}=17.8$ and $r_{\mathrm{ds}}=32.1$ for $r_{\mathrm{in}}=6$ and $r_{\mathrm{out}}=45$. 
The impulse response is set by $t_{0}=4.3\,\si{ms}$, $\Delta t_{0}=5.0\,\si{ms}$, where the reprocessing starts at $t_{0}-\Delta t_{0}/2 = 1.8\,\si{ms}$ corresponding to the light crossing of $45 R_{\mathrm{g}}$ for $M=8M_{\odot}$. 
We also show the effect of
changing $t_{0}$ on the lag-frequency spectra 
in Fig.~\ref{fig:psd_lagf_comp_discontinuous_fvisc}, with 
$t_{0}=2.8\,\si{ms}$ (magenta) and $6.1\,\si{ms}$ (orange). 
Plainly the observed switch between propagation and reverberation at $\sim 3$~Hz actually favours longer time lags rather than shorter ones, but these exacerbate the discrepancy below 3~Hz. 
All other parameter values are the same as in Fig.~\ref{fig:psd_comp_discontinuous_fvisc}.

We check the consistency of the derived impulse response with the derived accretion flow geometry because the top-hat impulse response is one of the simplest assumptions of the reflection/reprocessing.
In reality, the impulse response is affected by many factors, such as the geometry of the illuminating source to the disc and the inclination angle of the accretion disc to the observer.
The derived light crossing of $45R_{\mathrm{g}}$ does not mean the disc truncation of $45R_{\mathrm{g}}$, even if it is assumed that the reflection/reprocessing happens at the truncated disc.

Fig.~\ref{fig:timeprop_diffgeo} shows the comparison of impulse responses (a), power spectra in the low energy band (b), and lag frequency spectra (c) for different formulations of the impulse response. 
The black lines represent the case of the top-hat impulse response adopted above.
The red and blue solid lines represent the more realistic treatments of the lamp-post and extended illuminating source geometries, respectively.
In the lamp-post geometry, a source height of $h=10$ in units of $R_{\mathrm{g}}$ is assumed, while in the extended geometry, we consider a cylinder with the inner radius of $r_{\mathrm{in}}=6$, the outer radius of $r_{\mathrm{out}}=45$, and the height of $h=36$.
The reflection/reprocessing is assumed to happen at the outer thin disc from $r_{\mathrm{out}}=45$ to $r_{\mathrm{disc}}=400$, and contributions from further radii to the flux are ignored due to the small illuminating flux from the primary source (\citealt{Mahmoud_2019, Chainakun_2021}).
The magenta dashed line is calculated for the same lamp-post source geometry as in the red line ($h=6$), but the thin disc is assumed to be untruncated ($r_{\mathrm{out}}=6$), where the reflection/reprocessing occurs from $r_{\mathrm{out}}=6$ to $r_{\mathrm{disc}}=400$.
The inclination angle of $66^{\circ}$ is used, as in the spectral fitting.
Details of the calculation of the more realistic impulse responses are described in Appendix~\ref{sec:app_e}.
Note that the model is not self-consistent for the red solid line and magenta dashed line as it assumes both the radially extended hot flow and the disc truncation of $45R_{\mathrm{g}}$ in calculating the propagating fluctuations part.

The top-hat impulse response and that for the extended source geometry give similar power spectra and lag-frequency spectra, which demonstrates that the simple top-hat impulse response characterised by $(t_{0}, \Delta t_0)=(4.3\,\si{ms}, 5.0\,\si{ms})$ is a reasonable approximation to more realistic impulse responses for truncated radii of $45R_{\mathrm{g}}$. The mean reverberation delay time for the best fit top-hat is 4.3~ms, whereas that for the self-consistent extended source/truncated disc with $r_{\mathrm{out}}=45$ is very similar at 4.9~ms. This
is predominantly set by $r_{\mathrm{out}}$ rather than the emissivity as a lamp-post at $h=10$ gives a very similar mean delay of 4.7~ms for $r_{\mathrm{out}}=45$, whereas the delay reduces to 1.8~ms for a lamp-post at $h=10$ irradiating a disc with $r_{\mathrm{out}}=6$. 
The untruncated disc then gives shorter soft lags, so these
dominate over the propagation lag at  
higher frequencies than for the truncated disc models.  
Thus, it is more difficult to explain the observed soft lag with the untruncated disc geometry in our model.

However, there are still some limitations to our model. It underestimates the propagation lags in the $0.3$--$3\,\si{Hz}$ frequency range. 
More fundamentally, the power spectra do not reproduce the shift in peak frequency of $\mathrm{P}_{2}$ with energy since the reprocessed disc variability now matches the higher-energy power spectra rather than being shaped by propagation. These 
issues are discussed further in Section~\ref{sec:8_2}.


\section{Discussion}
\label{sec:8}
Our model produces the double-humped power spectrum by assuming there are intrinsic low-frequency fluctuations on the inner edge of the truncated disc, with a distinct jump to higher-frequency fluctuations in the hot flow. 
The turbulent disc region may be quite limited in radial extent, making only a single Lorentzian, $\mathrm{P}_{1}$, whereas the hot flow spans a larger range in radii, so the intrinsic fluctuations produce Lorentzians of different frequencies. 
Propagation of the fluctuations means that the power spectrum at any radius carries the imprint of variability generated at all larger radii. 
Thus at the inner edge of the disc, the power spectrum is only $\mathrm{P}_{1}$, whereas at any radius, $r$, in the hot flow it is the sum over all radii in the hot flow of the much higher-frequency Lorentzians at each radius, plus $\mathrm{P}_{1}$ from the truncation region. 

We couple this radial stratification of variability with radial stratification of the spectral components: the disc is (a sum of) blackbodies, whereas the Comptonisation in the hot flow is softer closer to the disc and harder closer to the central object. 
The combination of the radial stratification predicts that the blackbody has power spectrum $\mathrm{P}_{1}$, the soft Comptonisation has power spectrum which is $\mathrm{P}_{1}$ plus the sum of Lorentzians from the outer radius of the hot flow, $r_{\mathrm{ds}}$ down to $r_{\mathrm{sh}}$, and the hard Comptonisation power spectrum is this plus the additional noise generated from $r_{\mathrm{sh}}$--$r_{\mathrm{in}}$. 
This picture produces the observed asymmetric shape (broader than a single Lorentizan) of the second hump in the power spectrum, $\mathrm{P}_{2}$, and its increasing high-frequency extent with increasing energy due to the increasing contribution from the hard Comptonisation. 
The hot flow illuminates the disc, producing a fast variable reprocessed thermal component which adds to the intrinsic, slowly variable disc emission.

The power spectrum in any energy band is then determined by the contribution of each spectral component in each energy band. 
Similarly, the lag between correlated variability at a given frequency in any two energy bands will also be determined by a combination of propagation and reverberation lags in each spectral component and the changing contribution of the different spectral components with energy. 
Our model has a light travel time to the reverberating disc of 
$\sim 45R_{\mathrm{g}}$, as required in the truncated disc models and consistent with the inferred intrinsic thermal component. 
This combined spectral-timing model fits the time-averaged 
energy spectrum (by construction) but can also reproduce most of the features of the power spectra as a function of energy and the lags as a function of frequency. 

Our timing model can be used to test hot flow models in terms of the the radial velocity (accretion speed) $v_{r}(r)$ or equivalently, the viscous time-scale (accretion time-scale) $t_{\mathrm{visc}} (r)$ (see equation (\ref{eq:radial_velocity})).
We compare our results to various hot flow models, but stress that our timing model probably does not reproduce the observed properties sufficiently 
well to reach a robust conclusion.

\subsection{Linking the viscous time-scale to accretion flow models}
\label{sec:8_1}
\begin{figure}
	\includegraphics[width=\columnwidth]{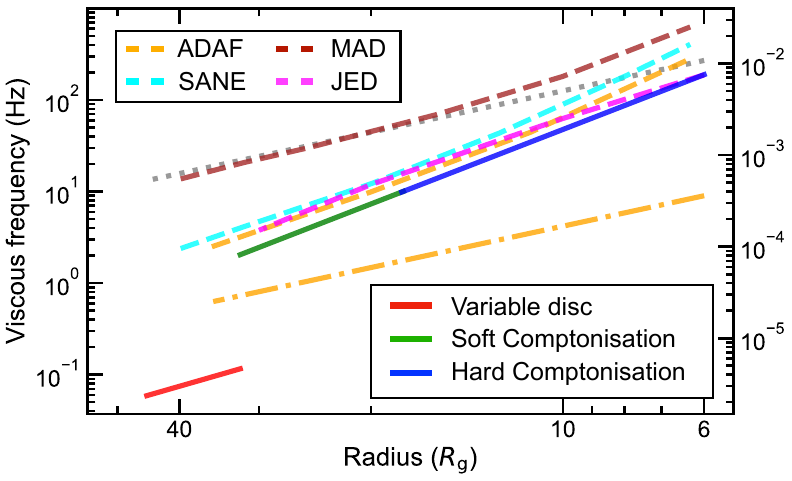}
    \caption{Viscous frequency as a function of radius from the central object. 
    Whereas the left axis shows it in units of $\si{Hz}$, the right axis in units of $c/R_{\mathrm{g}}$.
    The black hole mass of $M=8M_{\odot}$ is assumed.
    The solid lines show the measure from our model (see Fig. \ref{fig:psd_lagf_comp_discontinuous_fvisc}), and red, green, and blue colours correspond respectively to the disc, soft Comptonisation, and hard Comptonisation regions.
    The orange lines show the numerical (dashed) and self similar (dash-dotted) solutions from the ADAF model with $\alpha=0.01$ (\citealt{Narayan_1997}).
    The prediction from the SANE and MAD models (chunk S6 and M5 in \citealt{Narayan_2012} respectively) is shown with the cyan and brown dashed lines. 
    The prediction from the JED-SAD model with the Mach number, defined as the accretion speed to sound speed ratio, being unity is shown with the magenta dashed line (\citealt{Marcel_2018a, Marcel_2018b}).
    The Keplerian frequency $f_{\mathrm{K}}(r)$ is shown with the grey dotted line for a reference.}
	\label{fig:viscous_freq_radius}
\end{figure}

The comparison of viscous frequency as a function of radius derived from the timing model and that of various hot flow models are shown in Fig. \ref{fig:viscous_freq_radius}. Generically, these have 
radial velocity $v_{r}=\alpha (H/R)^2 v_{\phi} (R)$, where $\alpha$, $R$, $H$, and $v_{\phi} (R)$ are the viscosity parameter, radius, scale height, and rotation velocity of the flow, respectively (\citealt{Kato_2008}). Thus the
purely analytic, self similar Advection Dominated Accretion Flow (ADAF) solutions of \cite{Narayan_1994}, which have constant $\alpha$ and $H/R$, predict 
$v_{r}\propto v_{\phi}\propto r^{-1/2}$
(orange dash-dotted line), whereas our derived model has a much steeper dependence. 
However, including the stress-free inner boundary condition in a global ADAF model (orange dashed line) 
recover a good match for $\alpha\sim 0.01$--$0.03$ (\citealt{Narayan_1997, Popham_1998}) as the flow must accelerate inwards.

We also look at global MRI simulations of hot flows, where the angular momentum transport is self-consistently calculated by the magneto-rotational instability (MRI), so $\alpha$ is a function of radius.
The Standard And Normal Evolution' (SANE) model of \cite{Narayan_2012} gives viscous time-scales (cyan dashed line in Fig. \ref{fig:viscous_freq_radius}) which match well to our derived viscous time-scales. 
Other SANE models (\citealt{Bollimpalli_2020}) also show similar viscous time-scales to those derived from \cite{Narayan_2012} and thus match our spectral-timing model.

However, recent simulations have focused instead on Magnetically Arrested Discs (MAD; \citealt{Narayan_2003, Narayan_2012}). 
These have maximal magnetic flux pinned onto the black hole horizon, 
such that the magnetic field is strong enough that its pressure exceeds the ram pressure of the infalling material. 
In 1D calculations, this halts the accretion, but full 3D calculations 
show that accretion takes place instead via the magnetic interchange instability (\citealt{White_2019, Xie_2019}). 
The motivation for such states is that they power strong jets. 
MAD models have faster radial velocity than the SANE models as the large-scale magnetic fields give more efficient angular momentum transport than the small-scale dynamo (\citealt{Narayan_2012}). 
However, this means that they predict an accretion speed which is too fast to match our derived constraints (brown dashed lines in Fig.~\ref{fig:viscous_freq_radius}).

The jet can also be incorporated in a rather different way, by assuming a large scale height magnetic field exists rather than calculating it {\it ab initio} (the Jet Emitting Disc models (JED); \citealt{Marcel_2018a}). 
If this field is responsible for transporting angular momentum in the hot inner flow then the effective $\alpha\sim 3$. 
However, this can still give a good match to our velocity profile (magenta dashed line in Fig. \ref{fig:viscous_freq_radius}) as these models have a rather small scale height such that $H/R \sim 0.1$ contrasting with the larger scale height global ADAF, SANE and MAD flows. 

However, while the SANE and JED models both give a good match to our derived radial velocity, there is still a potential observable difference between them in their predicted sound speed, $c_{\mathrm{s}} \sim (H/R) v_{\phi}(R)$ (\citealt{Kato_2008}). 
The SANE models have $\alpha < H/R$, and thus $c_{\mathrm{s}}>v_{r}$, so the sound waves called the bending waves can cross the flow before they are damped out by viscosity. 
The flow can then globally precess if it is misaligned with the black hole spin (the wave-like regime; e.g. \citealt{Ingram_Motta_2019}). 
This is the most compelling current model for the origin of the low-frequency QPO (\citealt{Ingram_2009}). 
By contrast, the JED/MAD models have $\alpha > H/R$, and thus $c_{\mathrm{s}}<v_{r}$, so viscosity damps out any misalignment torques into a stable warp, which does not precess (the diffusive regime), ruling out Lense-Thirring precession as the origin of the QPO in the JED models 
(\citealt{Marcel_2021}). If global precession really switches off sharply at $\alpha = H/R$ then a Lense-Thirring origin of the QPO favours the SANE models. If instead there is a more gradual transition at $\alpha=H/R$ (so that precession is damped but still possible) then the JED models could give an origin for both the QPO and the jet.

\subsection{Limitations of the variability model}
\label{sec:8_2}
The timing model predicts time lags significantly lower than that observed for $0.3$--$3\,\si{Hz}$, 
and overestimates the frequency at which the soft (negative) time lag reaches its maximum (Fig.~\ref{fig:psd_lagf_comp_discontinuous_fvisc}, lower panel).
It is difficult to solve these discrepancies within the current spectral decomposition (fixed to that in Fig.~\ref{fig:ni1200120106_eeuf_rat}).
There could be a better match to the timing properties from a different set of spectral components.
We fixed the emissivity of the whole variable flow, which remains unclear, to that of the standard disc throughout the paper.
Different emissivity yields different timing properties, although it could be uncomfortably fine-tuned (\citealt{Ingram_2011, Ingram_2012}).
However, we note that the different emissivity does not significantly change the truncation radius in our model since it is derived from the low-frequency break in the power spectra, which simply marks the viscous frequency at the truncation radius (\citealt{Ingram_2011}).
The extensive exploration of the spectral-timing analysis is beyond the scope of the paper.

Our final model also does not reproduce the observed shift in peak frequency of $\mathrm{P}_{2}$ in power spectra, as shown in Fig.~\ref{fig:psd_lagf_comp_discontinuous_fvisc} (upper panel).
This is a qualitative discrepancy between the model and observation, and not solved as long as we simply attribute $\mathrm{P}_{1}$ and $\mathrm{P}_{2}$ to the variable disc and hot flow, respectively.
In this regard, we conclude that our current model lacks some fundamental factor(s) that impacts the fast variability.
The discrepancies are seen particularly at low energies where the disc emission significantly contributes to the spectrum, and mid frequencies between $\mathrm{P}_{1}$ and $\mathrm{P}_{2}$.
This suggests a complicated interplay between the truncated disc and hot flow at their transition region.
The complete discontinuity of viscous time-scale at the transition radius $r_{\mathrm{ds}}$ may be an oversimplification.
Due to the variability of the seed photons supplied by the inner edge of the truncated disc, the spectral shape of Comptonisation components is likely to vary in time (\citealt{Kotov_2001, Veledina_2016, Mastroserio_2018, Veledina_2018}), which is not taken into account in the current model.

We show here a proof of concept that the variability data allow us to probe the dynamics of the X-ray emitting region in such a way as to enable us to distinguish the nature and geometry of the hot flow, and how it connects to the jet.
We plan to improve the timing model and perform a more comprehensive spectral-timing analysis in future works.


\section{Conclusions}

We have performed a full spectral-timing analysis for MAXI J1820+070 in the low/hard state. This is based on the spectral components identified in the very detailed {\it NuSTAR} study of \cite{Zdziarski_2021b}, though we show the data favour an additional narrow iron line which is 
likely contributed from the strong wind seen in optical spectra in the state. We extended the model down to the {\it NICER} energy bandpass, including thermal emission from the disc. We see evidence for atomic features in this component also, as expected from the low-temperature disc photosphere at $\sim 0.22$~keV. 

Analysing the fast variability, we find the standard double-humped power spectrum seen in this state, indicating two variability components related to different time-scales, $\mathrm{P}_{1}$ and $\mathrm{P}_{2}$, which behave differently with respect to energy: whereas the component $\mathrm{P}_{1}$ maintains both its shape and amplitude for different energies, the peak frequency shifts and the high-frequency variability increases with energy in the component $\mathrm{P}_{2}$. 
The {\it NICER} data allow us for the first time to investigate this behaviour with good statistics below $2\,\si{keV}$. 

To explore the observed timing properties, we have developed a timing model based on the propagating fluctuations of the mass accretion rate.
The spectral decomposition derived from the fit is used to define the energy spectra emitted at each radius in the flow.
We have developed a full spectral-timing model which incorporates discontinuity of the viscous frequency at the transition radius between the truncated disc and hot flow, as well as a spectrally inhomogeneous hot flow. 
The discontinuity of the viscous frequency naturally reproduces the double-humped shape corresponding to $\mathrm{P}_{1}$ and $\mathrm{P}_{2}$ in the power spectra, as shown by \cite{Rapisarda_2016}, but our full spectral-timing model also allows us to explore the observed energy dependence of $\mathrm{P}_{2}$. 
We find that this can be explained by propagating fluctuations through the radially stratified (soft and hard Comptonisation components) hot flow. 

We also implement reflection/reprocessing of the Compton components on the disc.
Using the disc truncation at a few tens of gravitational radii, we show that this can give a good overall match to the lag-frequency spectrum, with the switch between soft leads (propagation) at low frequencies and soft lags (reverberation) at high frequencies, as well as fitting to the shape of the energy-dependent power spectra.
Our results on the reverberation size scale are consistent with our assumed truncated disc/hot inner flow geometry. 
While our timing model does not fit all aspects of the data, we show that reverberation from a much smaller inner disc radius (as derived from some iron line fits: \citealt{Buisson_2019}) makes these discrepancies worse. 

The viscous time-scale constrains the combination of effective viscosity and scale height of the flow. 
The SANE models (\citealt{Narayan_2012}), where the MRI dynamo transports the angular momentum, have fairly low effective viscosity and large scale height, and these match well to our constraints. 
By contrast, the MAD models (\citealt{Narayan_2012}) have similar scale heights, but higher effective viscosity from the large scale magnetic fields, so predict viscous time-scales which are substantially faster than those observed. 
Nonetheless, the JED models (\citealt{Marcel_2018a}) with large scale magnetic fields linked to the jet can match our derived viscous time-scales, where a smaller scale height flow offset the higher efficiency of the large scale magnetic torques. 
However, these have a sound speed which is slower than the viscous speed. 
This is important as Lense-Thirring precession, currently the best model for the low-frequency QPO (\citealt{Ingram_2009}), depends on 
misalignment torques transmitted by bending waves. 
These travel at $\approx$ the sound speed, but are damped out if the viscous speed is faster than the sound speed (\citealt{Ingram_Motta_2019}). If this really is able to completely suppress the global precession \citep{Marcel_2021} then we favour models where the angular momentum transport in the hot flow is not caused by a large-scale, ordered magnetic field linked to the jet, but instead is powered by the small-scale MRI dynamo. If instead there is more complex behaviour around the transition from the wave-like to diffusive regime, then models which also tie the jet into the hot flow are clearly attractive. 

The timing model still has room for improvement, particularly for the low energy bands, where the disc emission dominates the energy spectrum. 
Nonetheless, this study demonstrates the potential for spectral-timing models to address the fundamental nature and geometry of the accretion flow close to the black hole, and its link to the jet. 


\section*{Acknowledgements}
We thank G. Marcel and P. O. Petrucci for illuminating discussions about the JED models. 
We acknowledge valuable comments from the referee that helped to improve the manuscript.
This work was supported by JSPS KAKENHI Grant Numbers 18H05463 and 20H00153 and World Premier International Research Center Initiative (WPI), MEXT, Japan. 
TK acknowledges support from JSPS Overseas Challenge Program for Young Researchers and JST SPRING, Grant Number JPMJST2108.
CD acknowledges support from STFC through grant ST/P000541/1,
and Kavli Institute for the Physics
and Mathematics of the Universe (IPMU) funding from the National
Science Foundation (No. NSF PHY17-48958).


\section*{Data availability}
This work has made use of data and software provided by the High Energy Astrophysics Science Archive Research Center (HEASARC) Online Service managed by the NASA/GSFC. 


\bibliographystyle{mnras}
\bibliography{manuscript.bib} 



\appendix


\section{Model calculation of power spectra and time lags}
\label{sec:app_a}

We summarise the practical method to calculate the power spectra and frequency-dependent time lags of the flux in our model.
The method is mainly based on \cite{Ingram_2013}.
First, we take into account the inwards propagation of mass accretion rate fluctuations alone.
The case, where the reflection by the disc is included, is described in the following appendices.
Since the observation data are discrete time series, we should take discrete times $t_{j}$ and frequencies $f_{m}$ with integer subscripts $j$, $m$, rather than continuous $t$, $f$, in the formulation.
However, in some contexts, we use the notation as the continuous time and frequency in order to simplify expressions.

Under the inwards propagation of the fluctuations in a multiplicative manner (\citealt{Uttley_2005}), the dimensionless local mass accretion rate at the $n$th ring ($n=1, \cdots, N_{\mathrm{r}}$) follows 
\begin{equation}
    \dot{m}(r_{n}, t) = \prod _{k=1} ^{n} a(r_{k}, t-\Delta t _{k, n}),
\end{equation}
where $a(r_{k}, t)$ is the intrinsic fluctuations generated at the $k$th ring and $\Delta t_{k, l}$ is the propagation time from the outer $k$th ring to the inner $n$th ring ($k\leq n$).
From (\ref{eq:radial_velocity}) the propagation time is written as 
\begin{equation}
    \Delta t_{k, n} = \frac{\Delta r}{r} \sum _{l=k+1} ^{n} \frac{1}{f_{\mathrm{visc}}(r_{l})}.
\end{equation}
The ratio $\Delta r/r := \Delta r_{n}/r_{n}$ is constant for any rings ($n=1, \cdots, N_{\mathrm{r}}$) due to the logarithmic spacing of them.

According to \cite{Ingram_2013}, the modulus squared of the Fourier transform of the local mass accretion rate at $n$th ring is given by 
\begin{equation}
    |\dot{M}(r_{n}, f)|^2 \propto \coprod _{k=1} ^{n} P(r_{k}, f), 
    \label{eq:psd_mdot}
\end{equation}
where $\dot{M}(r_{n}, f)$ is the Fourier transform of the local mass accretion rate at $n$th ring and $P(r_{k}, f)$ the power spectrum of the intrinsic fluctuations of $k$th ring (\ref{eq:mdot_psd_intrinsic}).
The symbol of the co-product, $\coprod$, denotes the successive convolutions in frequency, which can be calculated efficiently by combining the convolution theorem with fast Fourier transforms (\citealt{Press_1992}).
The product between the complex conjugate of the Fourier transform of the local mass accretion rate at an outer $k$th ring and the Fourier transform at an inner $n$th ring ($k<n$) is 
\begin{equation}
    (\dot{M}(r_{k}, f))^{*} \dot{M}(r_{n}, f) \propto \Lambda_{k, n} \mathrm{e}^{-i 2\pi f \Delta t_{k, n}} P(r_{k}, f),
    \label{eq:csd_mdot}
\end{equation}
in which $*$ denotes the complex conjugate and $i$ the imaginary unit.
The product of the average of the local mass accretion rate $\mu_{l}$ ($l=1, \cdots, N_{\mathrm{r}}$)from $k$th ring to $n$th ring is expressed by $\Lambda _{k, n}$: $\Lambda _{k, n}=\prod _{l=k+1} ^{n} \mu_{l}$.
Since we assume that the average of the local mass accretion rate is unity for all rings, $\mu_{l}=1\,(l=1, \cdots, N_{\mathrm{r}})$, the product $\Lambda _{k, n}$ is always reduced to unity.
The power spectrum and cross spectrum of the local mass accretion rate can be obtained respectively by giving appropriate normalisation to equations (\ref{eq:psd_mdot}) and (\ref{eq:csd_mdot}).

In the derivation of equation (\ref{eq:csd_mdot}), we define the discrete Fourier transform of time series $y_{j} (j=0, 1, \cdots, N-1)$ sampled at the time $t_{j}=j\Delta t$ with the interval of $\Delta t$ during the time $T=N\Delta t$ as 
\begin{equation}
    a_{m} = \sum _{j=0} ^{N-1} y_{j} \mathrm{e}^{-i 2\pi mj / N } = \sum _{j=0} ^{N-1} y_{j} \mathrm{e}^{-i2\pi f_{m} t_{j}}.
\end{equation}
In case that the number of data $N$ is even, the subscript of the Fourier transform $a_{m}$ takes $m=-N/2+1, \cdots, N/2$, whereas in case $N$ is odd, it takes $m=-(N-1)/2, \cdots, (N-1)/2$.
The subscript is related to the frequency $f_{m}=m/T$.
The discrete inverse Fourier transform is then written as 
\begin{equation}
    y_{j}=\frac{1}{N} \sum _{m} a_{m} \mathrm{e}^{+i 2\pi mj / N } = \frac{1}{N} \sum _{m} a_{m} \mathrm{e}^{+i2\pi f_{m} t_{j}}.
\end{equation}

From equation (\ref{eq:flux_e_narrow}), the Fourier transform of the flux with energy $E$ is 
\begin{equation}
    X(E, f)=\sum _{n=1} ^{N_{\mathrm{r}}} w(r_{n}, E) \dot{M}(r_{n}, f).
    \label{eq:flux_fourier}
\end{equation}
Using equations (\ref{eq:csd_mdot}) and (\ref{eq:flux_fourier}), the modulus squared of $X(E, f)$ is written as 
\begin{equation}
    \begin{split}
    |X(E, f)|^2 \propto & \sum _{n=1} ^{N_{\mathrm{r}}} \biggl[ (w(r_{n}, E))^2 |\dot{M}(r_{n}, f)|^2 \\
                  & + 2 \sum _{k=1} ^{n-1} \Bigl( w(r_{k}, E) w(r_{n}, E) \Lambda _{k, n} \cos (2\pi f \Delta t_{k, n})\\
                  & \times  |\dot{M}(r_{k}, f)|^2 \Bigr) \biggr].
    \end{split}
    \label{eq:psd_flux}
\end{equation}
The product between the complex conjugate of the Fourier transform of the flux with the photon energy $E_{1}$ and the Fourier transform of the flux with the photon energy $E_{2}$ is 
\begin{equation}
    \begin{split}
    (X(E_{1}, f))^{*} X(E_{2}, f) \propto &\sum _{n=1} ^{N_{\mathrm{r}}} \biggl[ w(r_{n}, E_{1}) w(r_{n}, E_{2}) |\dot{M}(r_{n}, f)|^2 \\
                  & + \sum _{k=1} ^{n-1} \Bigl[ \Bigl( w(r_{k}, E_{1}) w(r_{n}, E_{2}) \mathrm{e}^{-i2\pi f \Delta t_{k, l}} \\
                  &+ w(r_{n}, E_{1}) w(r_{k}, E_{2}) \mathrm{e}^{+i2\pi f \Delta t_{k, l}} \Bigr) \\
                  & \times \Lambda _{k, n}  |\dot{M}(r_{k}, f)|^2 \Bigr] \biggr].
    \end{split}
    \label{eq:csd_flux}
\end{equation}
The power spectrum and cross spectrum of the flux can be calculated respectively by giving appropriate normalisation to equations (\ref{eq:psd_flux}) and (\ref{eq:csd_flux}).
The phase lag between two energies $E_{1}$ and $E_{2}$ as a function of frequency is expressed as 
$\phi _{\mathrm{lag}} (E_1, E_2, f)=\tan ^{-1} [\mathrm{Im}[(X(E_{1}, f))^{*} X(E_{2}, f)]/\mathrm{Re}[(X(E_{1}, f))^{*} X(E_{2}, f)]]$, 
where $\mathrm{Re}[\cdots]$ and $\mathrm{Im}[\cdots]$ denote the real and imaginary parts, respectively.
The phase lag is converted into the time lag by 
$t_{\mathrm{lag}} (E_1, E_2, f) = \phi _{\mathrm{lag}} (E_1, E_2, f)/(2\pi f)$.
In the definition of the cross spectrum (\ref{eq:csd_flux}), the positive lag means that fluctuations at energy $E_{1}$ lag behind those at energy $E_{2}$.
The extension to a wide energy band $E_{\mathrm{min}} <E \leq E_{\mathrm{max}}$ can be performed by using equation (\ref{eq:weight_e_wide}) instead of equation (\ref{eq:weight_e_narrow}) as the weight.


\section{Model including reflection and reprocessing}
\label{sec:app_b}

First of all, taking the reprocessed emission into account changes the transition radii $r_{\mathrm{ds}}$, $r_{\mathrm{sh}}$ because the total flux of the intrinsic disc emission decreases to $(1-f_{\mathrm{rep}})S_{\mathrm{d}}(E)$ and so in the relation used to derive the transition radii (\ref{eq:trans_radii}), $S_{\mathrm{d}}(E)$ is replaced by $(1-f_{\mathrm{rep}})S_{\mathrm{d}} (E)$.
We summarise alterations in the variability of the flux below.

To simplify the formulation, we treat the flux designated by a single photon energy $E$.
When the reflection is taken into account, the flux is expressed as 
\begin{equation}
    x(E, t)=x_{\mathrm{dir}}(E, t) + x_{\mathrm{ref}}(E, t),
\end{equation}
where $x_{\mathrm{dir}}(E, t)$ is the direct component from the variable flow and $x_{\mathrm{ref}}(E, t)$ reflected component resulting from the response of the disc to the illumination by the Comptonisation region.

The former component $x_{\mathrm{dir}}(E, t)$ is written as the form of equation (\ref{eq:flux_e_narrow}) with the slight modification of the weight (\ref{eq:weight_e_narrow}), that is $S_{\mathrm{d}}(E)$ is replaced by $(1-f_{\mathrm{rep}})S_{\mathrm{d}}(E)$.
We define the modified weight as $w_{\mathrm{dir}}(r_{\mathrm{n}}, E)$.

The latter component $x_{\mathrm{ref}}(E, t)$ is written as 
\begin{equation}
    x_{\mathrm{ref}}(E, t)=h(E, t) \otimes x_{\mathrm{f}}(t), 
\label{eq:flux_reprocessing}
\end{equation}
where $h(E, t)$ is the impulse response of the reflection (\ref{eq:impulse_response}) and $x_{\mathrm{f}}(t)$ the flux coming from the whole Comptonisation region.
The symbol $\otimes$ denotes the convolution.
The flux $x_{\mathrm{f}}(t)$ can be expressed as the form of equation (\ref{eq:flux_e_narrow}) with the following weight:
\begin{equation}
    w_{\mathrm{f}}(r_{n})=
    \begin{cases}
        \int dE w(r_n, E) & (r_{\mathrm{in}} < r_{n} \leq r_{\mathrm{ds}}),\\
        0                 & (\text{otherwise}).
    \end{cases}
\label{eq:weight_compton}
\end{equation}
Note that the flux $x_{\mathrm{f}}(t)$ is treated as the energy-integrated form since we assume that the impulse response does not depend on the input photon energy.

The Fourier transform of the flux $x(E, t)$ is 
\begin{equation}
    X(E, f)=X_{\mathrm{dir}}(E, f)+H(E, f) X_{\mathrm{f}}(f),
\end{equation}
where the capital letters denotes the Fourier transform of the small letters.
The Fourier transform of the impulse response, $H(E, f)$, is called the transfer function and expressed as 
\begin{equation}
    H(E, f)=C(E)\Delta t_0 \mathrm{e}^{-i 2 \pi f t_0} \mathrm{sinc} \left( \pi f \Delta t_{0} \right).
\label{eq:transfer_function}
\end{equation}
Then the modulus squared of $X(E, f)$ is 
\begin{equation}
\begin{split}
    |X(E, f)|^2 = & |X_{\mathrm{dir}}(E, f)|^2 + |X_{\mathrm{f}}(f)|^2 |H(E, f)|^2 \\
                  & + 2\mathrm{Re}\left[ H(E, f) (X_{\mathrm{dir}}(E, f))^{*} X_{\mathrm{f}}(f) \right]. 
\end{split}
\label{eq:psd_reprocessing}
\end{equation}
Since both $x_{\mathrm{dir}}(E, t)$ and $x_{\mathrm{f}}(t)$ are expressed in the form of (\ref{eq:flux_e_narrow}), each term in equation (\ref{eq:psd_reprocessing}) can be calculated analytically, based on equations (\ref{eq:psd_flux}) and (\ref{eq:csd_flux}).
The product of the complex conjugate of the Fourier transform of the flux with the photon energy $E_{1}$ and the Fourier transform with another photon energy $E_{2}$ is 
\begin{equation}
\begin{split}
    (X(E_{1}, f))^{*} X(E_{2}, f) = & (X_{\mathrm{dir}}(E_1, f))^{*} X_{\mathrm{dir}}(E_2, f) \\
                                    & + (H(E_1, f))^{*} H(E_2, f) |X_{\mathrm{f}}(f)|^2 \\
                                    & + (H(E_1, f))^{*} (X_{\mathrm{f}}(f))^{*} X_{\mathrm{dir}}(E_2, f) \\
                                    & + H(E_2, f) (X_{\mathrm{dir}}(E_1, f))^{*} X_{\mathrm{f}}(f).
\end{split}
\label{eq:csd_reprocessing}
\end{equation}
For the same reason above, each term in equation (\ref{eq:csd_reprocessing}) can be calculated analytically.
Giving appropriate normalisation to equations (\ref{eq:psd_reprocessing}) and (\ref{eq:csd_reprocessing}) yields the power spectrum and cross spectrum of the flux.

The formulation described above is extended to the case, where a wide energy band $E_{\mathrm{min}} <E \leq E_{\mathrm{max}}$ is taken.
In the case, the weight for the direct component $w_{\mathrm{dir}}(E, t)$ is modified into $w_{\mathrm{dir}}((E_{\mathrm{min}}, E_{\mathrm{max}}), t)$, which is obtained from equation (\ref{eq:weight_e_wide}) with the substitution of  $(1-f_{\mathrm{rep}})S_{\mathrm{d}}(E)$ for $S_{\mathrm{d}}(E)$. 
The calculation of the normalisation of the impulse response is described in Appendix~\ref{sec:app_c}.
It is noted that $S(E)$ in equation (\ref{eq:weight_e_narrow}) should be replaced by $S(E) A_{\mathrm{eff}} (E) \mathrm{e}^{-N_{\mathrm{H}}(E)\sigma _{\mathrm{T}}}$ to take the different statistics for different energy bins in observations into account.


\section{Amplitude of the impulse response of the reprocessing}
\label{sec:app_c}
From equations (\ref{eq:flux_e_narrow}), (\ref{eq:impulse_response}), (\ref{eq:flux_reprocessing}), the time-averaged flux of the reflected emission with a photon energy $E$ is 
\begin{equation}
    \langle x_{\mathrm{ref}}(E, t) \rangle = C(E)\Delta t_0 \sum _{n=1} ^{N_{\mathrm{r}}} w_{\mathrm{f}}(r_{n}), 
\end{equation}
where the average of the local mass accretion rate, $\mu=1$, is used.
Since it is also expressed as $\langle x_{\mathrm{ref}}(E, t) \rangle = f_{\mathrm{rep}}S_{\mathrm{d}}(E)+S_{\mathrm{sr}}(E)+S_{\mathrm{hr}}(E)$, 
the amplitude of the impulse response is 
\begin{equation}
    C(E) = \frac{f_{\mathrm{rep}}S_{\mathrm{d}}(E)+S_{\mathrm{sr}}(E)+S_{\mathrm{hr}}(E)}{\Delta t_0 \sum _{n=1} ^{N_{\mathrm{r}}} w_{\mathrm{f}}(r_{n})}.
\label{eq:norm_impulse_response_e_narrow}
\end{equation}
When a wider energy band $E_{\mathrm{min}} <E \leq E_{\mathrm{max}}$ is taken, the amplitude $C(E)$ is modified into 
\begin{equation}
\begin{split}
    &C(E_{\mathrm{min}}, E_{\mathrm{max}})\\
    &= \frac{f_{\mathrm{rep}}S_{\mathrm{d}}(E_{\mathrm{min}}, E_{\mathrm{max}}) +S_{\mathrm{sr}} (E_{\mathrm{min}}, E_{\mathrm{max}}) +S_{\mathrm{hr}} (E_{\mathrm{min}}, E_{\mathrm{max}}) }{\Delta t_0 \sum _{n=1} ^{N_{\mathrm{r}}} w_{\mathrm{f}}(r_{n})}.
\end{split}
\label{eq:norm_impulse_response_e_wide}
\end{equation}
Again, $S(E)$ in equations (\ref{eq:weight_e_narrow}) and  (\ref{eq:average_flux}) should be multiplied by $A_{\mathrm{eff}} (E) \mathrm{e}^{-N_{\mathrm{H}}(E)\sigma _{\mathrm{T}}}$ to take the different statistics for different energy bins in observations into account.


\section{Comparison of timing properties with different parameter values}
\label{sec:app_d}

\begin{figure*}
	\includegraphics[width=0.9\linewidth]{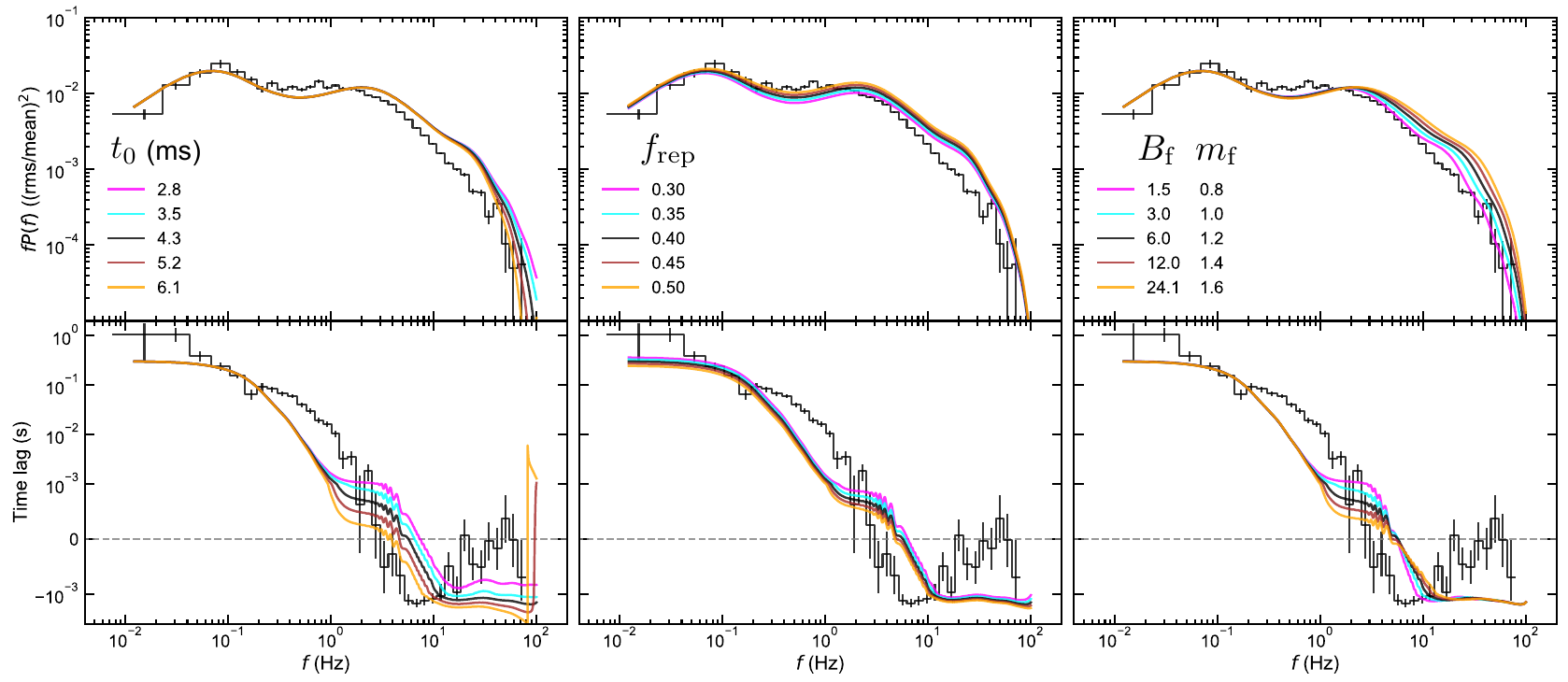}
    \caption{Power spectra at the low energy band of $0.5$--$1.5\,\si{keV}$ (upper) and lag-frequency spectra between $0.5$--$1.5\,\si{keV}$ and $2.0$--$10.0\,\si{keV}$ (lower) calculated from various values of model parameters, $t_{0}$ (left), $f_{\mathrm{rep}}$ (middle), and $(B_{\mathrm{f}}, m_{\mathrm{f}})$ (right). 
    The black lines are calculated with the parameter values tabulated in the bottom row of Table \ref{tab:model_par}. 
    Other lines are calculated by varying one or two parameter(s) with respect to the black lines. 
    Note that, although the time delay of the reflection $t_{0}$ and the fraction of the reprocessed emission $f_{\mathrm{rep}}$ should be related to the radii setting spectral regions in the accretion flow, $r_{\mathrm{sh}}$, $r_{\mathrm{ds}}$, and $r_{\mathrm{out}}$ (e.g. Appendix~\ref{sec:app_b}), these radii are fixed for simplicity.
    In the right column, the normalisation parameter of the viscous frequency in the hot flow, $B_{\mathrm{f}}$, is calculated such that the viscous frequency at the outer edge of the hot flow $r_{\mathrm{ds}}$ corresponds to the same value for different power indices $m_{\mathrm{f}}$.
    For comparison, the observation results are shown with the stepped lines.}
	\label{fig:discont_ref_psd_lagf_comp}
\end{figure*}

We show how model parameters affect the timing properties, picking the important parameters: the delay time of the reflection $t_{0}$, the fraction of the reprocessed emission $f_{\mathrm{rep}}$, and the power index of the viscous frequency in the hot flow $m_{\mathrm{f}}$.
As in Section \ref{sec:7_3}, the model including both the discontinuity of the viscous frequency and the reflection is used.
The power spectra at the low energy band of $0.5$--$1.5\,\si{keV}$ and lag-frequency spectra between $0.5$--$1.5\,\si{keV}$ and $2.0$--$10.0\,\si{keV}$ calculated with various parameter values are shown in Fig. \ref{fig:discont_ref_psd_lagf_comp}.
We explain the dependence of timing properties on each model parameter in order below.

From the left-top panel, the power spectrum is insensitive to the delay time $t_{0}$ because the power spectrum does not contain the information on the phases of the Fourier components, which is affected by $t_{0}$.
On the other hand, the lag-frequency spectrum is affected by $t_{0}$ at high frequencies, where the reflection contributes to the variability, as shown in the left-bottom panel.
The longer time delay increases the amplitude of the soft lag and makes the frequency at which the switch from the hard to soft lag occurs lower.

From the middle column, both the power spectrum and lag-frequency spectrum are generally insensitive to the fraction of the reprocessed emission $f_{\mathrm{rep}}$ if we use the same radii setting the spectral regions in the accretion flow.
The increase in $f_{\mathrm{rep}}$ slightly increases the variability of $\mathrm{P}_{2}$ in the low energy band, which is generated as a result of the reflection.
The increase in $f_{\mathrm{rep}}$ also slightly contributes to the shift of the time lags to the negative direction by decreasing the intrinsically variable disc emission resulting from the propagating mass accretion rate fluctuations.

The right column shows the dependence of timing properties on the power index parameter of the viscous frequency in the hot flow $m_{\mathrm{f}}$.
The normalisation $B_{\mathrm{f}}$ is calculated such that the viscous frequency at the outer edge of the hot flow, $\lim _{r \nearrow r_{\mathrm{ds}}}f_{\mathrm{visc}}(r)=B_{\mathrm{f}}r_{\mathrm{ds}}^{-m_{\mathrm{f}}} f_{\mathrm{K}}(r_{\mathrm{ds}})$ (see equation (\ref{eq:fvisc_discont})), corresponds to the same value for different $m_{\mathrm{f}}$.
The bigger power law index $m_{\mathrm{f}}$ means the larger variability at high frequencies, as shown in the right-top panel.
The lag-frequency spectra in the right-bottom panel show complicated dependence on $m_{\mathrm{f}}$.
For the larger $m_{\mathrm{f}}$, the transition from the hard to soft lag happens on a wider range of frequencies, although the frequency at which the time lag takes zero seems irrelevant to $m_{\mathrm{f}}$.

Fig. \ref{fig:discont_ref_psd_lagf_comp} also points to the difficulty in reproducing the timing properties related to low energy bands, where the disc contributes to the energy spectrum.
We plan to modify the timing model to explain the timing properties at these energy bands.


\section{Calculation of realistic impulse responses}
\label{sec:app_e}

We describe the calculation of impulse response of the outer disc for the X-ray illumination from the primary source with the lamp-post and extended geometries.
For simplicity, we assume that the outer disc is flat with infinitesimal thickness and optically thick and that the extended source is optically thin and radiation from everywhere in the source reaches the outer disc.
We formulate the case of the extended illuminating source first.
The impulse response for the lamp-post geometry can be calculated in the same way and more easily, as stated later.

We represent any points of the X-ray source and reflector with 
$\vb*{r}$ and $\vb*{r}'$. 
Adopting the cylindrical coordinate system with the origin being at the black hole and z-axis being perpendicular to the disc, we express them with
$\vb*{r} = (r\cos \phi, r\sin \phi, z)\,(r_{\mathrm{in}}\leq r < r_{\mathrm{out}}, 0 \leq \phi < 2\pi, -h\leq z < h)$ and 
$\vb*{r}' = (r'\cos \phi ', r' \sin \phi ', 0)\,(r_{\mathrm{out}} \leq r' < r_{\mathrm{disc}}, 0 \leq \phi < 2\pi)$, respectively.
The impulse response of the outer disc for the X-ray illumination is calculated with 
\begin{equation}
    h(t) \propto \int d \vb*{r} \int d \vb*{r}' \tilde{h}(\vb*{r}, \vb*{r}', t),
\label{eq:impulse_response_calc}
\end{equation}
where $\tilde{h}(\vb*{r}, \vb*{r}', t)$ is the local impulse response at the reflector $\vb*{r}'$ for the illuminating source $\vb*{r}$.
The local impulse response is expressed as a weighted delta-function:
\begin{equation}
    \tilde{h}(\vb*{r}, \vb*{r}', t) = \frac{\epsilon _{\mathrm{v}} (r) \cos (z/|\vb*{r}'-\vb*{r}|)}{|\vb*{r}'-\vb*{r}|^2} \delta (t-\tau (\vb*{r}, \vb*{r}')).
\label{eq:local_impulse_response}
\end{equation}
The volume emissivity of the hot flow is calculated by dividing the emissivity for surface by its height, $\epsilon _{\mathrm{v}}(\vb*{r})= \epsilon (r)/h$.
The terms, $|\vb*{r}'-\vb*{r}|^2$ and $\cos (z/|\vb*{r}'-\vb*{r}|^2)$, come from the inverse square law and correction of the effective area of the reflector, respectively (\citealt{Gardner_2017}).
The delay of the reprocessed emission with respect to the direct emission, $\tau (\vb*{r}, \vb*{r}')$, is written in units of $R_{\mathrm{g}}/c$ as (e.g. \citealt{Welsh_1991})
\begin{equation}
\begin{split}
    \tau (\vb*{r}, \vb*{r}') &= |\vb{r}'-\vb{r}|-(\vb{r}'-\vb{r}) \cdot \hat{\vb{r}}_{\mathrm{l}}\\
    &=\sqrt{(r' \cos \phi ' - r \cos \phi )^2 + (r' \sin \phi ' - r \sin \phi )^2 + z^2} \\
    & - \left( (r' \cos \phi ' - r \cos \phi ) \sin i - z\cos i \right),
\end{split}
\end{equation}
where we ignore any general relativistic corrections.
The inclination angle $i$ is measured between the axis of the disc and the line of sight.
Defining the x-axis and z-axis such that the line-of-sight is on $(x>0, y=0, z>0)$ plane without loss of generality, the unit vector of the line of sight is expressed as $\hat{\vb{r}}_{\mathrm{l}}$  $=(\sin i, 0, \cos i)$.
The integration for the illuminating X-ray source in equation (\ref{eq:impulse_response_calc}) is performed for $z>0$ since the source at $z<0$ illuminates the other side of the outer disc for the observer and so the reflected/reprocessed emission does not reach the observer.

For the lamp-post source geometry, the calculation of impulse response becomes simpler.
The integration for the source vanishes from equation (\ref{eq:impulse_response_calc}), and the consideration of emissivity of the source in equation (\ref{eq:local_impulse_response}) is no longer necessary.
Regardless of formulation, the amplitude of the impulse response $h(t)$ is calculated from the time-averaged spectrum of the reflected/reprocessed emission (see Appendix~\ref{sec:app_c} for the case of top-hat impulse response).


\bsp	
\label{lastpage}
\end{document}